\documentclass[12pt,preprint]{aastex}
\usepackage{graphics}
\usepackage{graphicx}
\usepackage{longtable,lscape}

\slugcomment{{\large Draft : \today}}

\begin{document}

\title{Radio Observations of the Hubble Deep Field South Region II: The 1.4 GHz Catalogue and Source Counts.}
\author{Minh T. Huynh\altaffilmark{1}} 
\affil{Research School of Astronomy \& Astrophysics, The Australian National
University, Mount Stromlo Observatory, Cotter Rd, Weston Creek, ACT 2611, Australia}
\email{mhuynh@mso.anu.edu.au}
\author{Carole A. Jackson} 
\affil{Australia Telescope National Facility, CSIRO Radiophysics Laboratory, PO Box 76, Epping, 
NSW 2121, Australia}
\author{Ray P. Norris}
\affil{Australia Telescope National Facility, CSIRO Radiophysics Laboratory, PO Box 76, Epping, 
NSW 2121, Australia}
\author{Isabella Prandoni}
\affil{Istituto di Radioastronomia, INAF, Via Gobetti 101, 40129, Bologna, Italy}

\altaffiltext{1}{affiliated with the Australia Telescope National Facility, CSIRO Radiophysics Laboratory, PO Box 76, 
Epping, NSW 2121, Australia}

\begin{abstract}
This paper is part of a series describing the results from the {\em Australia 
Telescope Hubble Deep Field South} (ATHDFS) survey obtained with the {\em Australia 
Telescope Compact Array} (ATCA). This survey consists of 
observations at 1.4, 2.5, 5.2 and 8.7 GHz, all centred on the Hubble Deep Field South.

Here we present the first results from the extended observing campaign at 1.4 GHz. 
A total of 466 sources have been catalogued to a local sensitivity of $5 
\sigma$ (11 $\mu$Jy rms). A source extraction
technique is developed which: 1) successfully excludes spurious sources from
the final source catalogues, and 2) accounts for the non-uniform noise in our
image. A source catalogue is presented and the general properties of the 1.4
GHz image are discussed. We also present source counts derived from our ATHDFS 1.4 GHz 
catalogue. Particular attention is made to ensure the counts are 
corrected for survey incompleteness and systematic effects. Our counts are
consistent with other surveys (e.g. ATESP, VIRMOS, and Phoenix Deep Field),
and we find, in common with these surveys, that the HDFN counts are systematically lower.

\end{abstract}


\section{Introduction}
Early radio surveys found that classical radio galaxies ($S \gtrsim 1$ Jy) 
evolve strongly with time in both density and luminosity, and are relatively 
rare in the local universe \citep{longair66}. More recent surveys which probe 
to fainter flux densities have shown that the normalised source count flattens
below a few mJy \citep{windhorst85}. This flattening corresponds to a rapid
increase in the number of faint radio sources, and has been interpreted as
being due to a new population of sources which do not show up at higher flux
density levels - the so called sub-mJy population. The existing sub-mJy 
samples are generally small due to the large amount of observing time 
required to reach these faint levels.

Several models, involving different classes of objects, have been developed to 
explain the observed sub-mJy source counts. These include a strongly evolving
population of normal spiral galaxies \citep{condon84, condon89} or 
actively star forming galaxies \citep{windhorst84, windhorst85,rowan93}. 
A non evolving population of local ($z < 0.1$) low 
luminosity radio galaxies has also been proposed to explain the sub-mJy 
population \citep{wall86}. It is now thought that the sub-mJy population is comprised of low
luminosity AGN, normal spirals and ellipticals, as well as starbursts, but the
exact mix is unknown. 

The nature of the sub-mJy population is still not clear because of the difficulty of obtaining optical 
photometry and spectra for complete radio samples. The follow up optical 
identification and spectroscopy of sub-mJy radio samples require a lot of telescope 
time because these sources generally have very faint optical counterparts. 
Conclusions about the sub-mJy sample have been inferred from surveys of
varying radio sensitivity and different depths of optical follow up.

Approximately 41\% of sub-mJy ($S_{\rm 1.4 GHz} > 0.2$ mJy) radio sources are identified with optical 
counterparts to $m_{B} \sim 23.7$ and these are mostly faint blue galaxies with 
optical-infrared colours indicative of active star formation
\citep{windhorst85}. This result is confirmed by
a spectroscopic study of optical counterparts ($B < 22$) to faint ($S_{\rm 1.4 GHz} >
0.1$mJy) radio sources, which found that most of the sub-mJy sources have 
spectra similar to star forming {\em IRAS} galaxies \citep{benn93}. However, 
this result is obtained from spectroscopy of only 10\% of their whole 
radio sample. 

A more spectroscopically complete sample is the Marano field, where
spectra were obtained for 50\% of the 68 faint ($ S_{\rm 1.4 GHz} > 0.2 $ mJy) radio sources 
\citep{gruppioni99}. Contrary to previous results 
of \cite{benn93}, the majority of the spectra are identified with 
early type galaxies. Gruppioni et al. attribute this 
discrepancy to the fainter magnitude limit reached in their spectroscopic 
identifications, since the fraction of radio sources identified with early 
type galaxies increases abruptly at $B \sim 22.5$, which is the limit of 
the Benn et al. sample. 

About 60\% (219/386) of the faint ($S_{\rm 1.4 GHz}> 0.5$ mJy) radio sources in the 
Australia Telescope ESO Slice Project (ATESP, \citealp{prandoni00a}) are
identified to I = 22.5 mag. The spectra obtained for 70 sources with I $<$ 19 
are dominated by early type galaxies (60\%). Starbursts and post-starburst 
galaxies become important in the sub-mJy ($S_{\rm 1.4 GHz} < 1$ mJy) regime, where 
they make up 39\% of the I $<$ 19 sample \citep{prandoni01b}. Nevertheless, 
Prandoni et al. find that early type galaxies still constitute a significant 
fraction (25\%) of the sub-mJy radio sources, and sub-mJy samples with 
fainter spectroscopic limits become increasingly sensitive to early type galaxies. 

The Phoenix Deep Field survey covers a 4.5 deg$^2$ at 1.4 GHz and has a rms noise of 12 $\mu$Jy over the most sensitive region \citep{hopkins03}. A subregion covering about a degree has been imaged to R $\sim$ 24.5 and optical counterparts found for 76\% of radio sources within the imaged area \citep{sullivan04}. Spectra of the radio sources revealed that the sub-mJy population is made up of at least 32\% starforming galaxies and 21\% are quiescent galaxies \citep{afonso05}. However, \cite{afonso05} only have spectra for 17\% of their radio sources, and the spectral completeness is biased to the optically brighter radio sources (R $\lesssim 21$). 

The most complete result in photometry is in the Hubble Deep Field North
(HDFN) where the VLA was used to follow up observations of a patch of sky observed by the HST to depths of
about 30th magnitude in the four HST broadband filters, F300W, F450W, F606W, and F814W 
(approximately U,B,V, and I). Eighty percent of the 79 faint ($S_{\rm 1.4 GHz} \geq 0.04$
mJy) radio sources in the HDFN region were identified by HST and
ground based images reaching $I_{\mbox{AB}} = 25$ \citep{richards98,richards00}. 
Using optical and radio morphologies along with radio spectral index arguments, Richards et al. 
concluded that 60\% of the HDFN radio sources with optical 
counterparts were star forming galaxies, 20\% were classed as AGN, and the remainder were ambiguous. 

In 1998, the HST observed a region of sky in the Southern continuous 
viewing zone (CVZ) as a complementary observation to the northern Hubble Deep Field 
\citep{williams96}. This field is known as the Hubble Deep Field South 
(HDFS) \citep{williams00}. A region containing a quasar suitable for the search of 
lyman alpha absorption systems was chosen. The main WFPC field of the HDFS 
observations reach a magnitude limit of $\sim$30th magnitude in
the four HST broadband filters. Simultaneous observations with the other HST 
instruments, NICMOS and STIS, reach similarly 
deep levels in the near-infrared and ultraviolet. In addition to the main HST deep 
fields, a mosaic of nine flanking fields were imaged to shallower levels of 
$\sim$25th in I (F814W). From the ground, deep wide field images have been obtained, 
reaching depths of $\sim$25th magnitude in UBVRI in a 44 $\times$ 44 arcmin region 
centered on the HDFS \citep{teplitz01}.  Spectroscopically, 194 galaxies in the 
main HDFS and Flanking Fields were targeted on the Very Large Telescope (VLT) 
and reliable redshifts obtained for 97 targets \citep{sawicki03}.
In addition, optical spectroscopy is available for 225 bright (R$ < 24$) 
sources in a $9 \times 3$ arcmin region containing the STIS and WFPC 
fields (Glazebrook, in preparation, but see {\tt http://www.aao.gov.au/hdfs/Redshifts/}).

The HDFS was selected as a region for deep radio follow up to take advantage 
of the wealth of publicly available optical/near-infrared photometric 
and spectroscopic data.
Observations over 1998 to 2001 were performed on the Australia Telescope Compact 
Array (ATCA) at all four available frequency bands. Approximately 100 hours observing 
at each band yielded images at 1.4, 2.5, 5.2, and 8.7 GHz with sensitivities of 
approximately 10 $\mu$Jy rms. A detailed description of observations, catalogues and 
analysis of these Australia Telescope Hubble Deep Field South (ATHDFS) images is 
described in Norris et al. (2004, hereafter Paper I). 

The radio source counts at the faintest flux density levels ($S_{\rm 1.4 GHz} < 1$
mJy) are not well determined due to the small number of surveys reaching 
this sensitivity regime. The sensitivity of our radio survey allows 
us to determine the source counts down to $\sim$0.05 mJy. 
Only the Phoenix Deep Field \citep{hopkins03} and
Hubble Deep Field North \cite{richards00} reach similarly deep levels. 

This paper describes the image analysis and source extraction technique of 
the full 1.4 GHz ATHDFS survey. This paper is organised as follows. In Section 
2 we summarise the 1.4 GHz observations and data reduction. 
Various image properties are discussed in Section 3 and the source 
extraction and catalogue is detailed in Section 4. Section 5 contains an 
analysis of the accuracy of our source parameters. The survey completeness is
investigated in Section 6 and the ATHDFS 1.4 GHz differential radio source
count is presented in Section 7.  

\section{Observations and Data Reduction}

Paper I discussed in detail the observations and data reduction steps used to obtain the
ATHDFS images. Here we provide a brief summary. 

The 1.4 GHz observations were carried out over four years from 1998 to
2001. They consist of single pointings 
centred on RA = 22h 33m 25.96s and Dec = $-$60$^\circ$ 38' 09.0'' (J2000). A wide
variety of ATCA configurations was employed to ensure maximum {\em uv}
coverage. The correlator was set to continuum mode ($2 \times 128$ MHz
bandwidth), with each 128 MHz bandwidth divided into $32 \times 4$ MHz channels.
The primary flux density calibrator used was PKS B1934-638, while 
secondary gain and phase calibrations were performed through observations 
of either PKS B2205-636 or PKS B2333-528. 

The data were inspected using the {\em MIRIAD} interactive task {\em
TVFLAG}. After flagging, the 1.4 GHz data were split into the three observing bands 
(1.344, 1.384 and 1.432 GHz) before imaging and cleaning. This allowed the
cleaning process to do a much better job of removing all the flux of a
source. Two iterations of both phase and amplitude self-calibration were then
performed to improve the image quality. 

\section{Image Analysis}
\label{imageanalysis}

In this Section we discuss two systematic effects which have to be taken into 
account in deriving source flux densities, the clean bias and bandwidth smearing. 
We use methods similar to those outlined in \cite{prandoni00b} to investigate
clean bias and bandwidth smearing.
A master equation to derive flux densities corrected for these effects is given 
in Appendix A.

\subsection{Clean Bias}
\label{cleanbias}

When the {\em uv} coverage is incomplete the cleaning process can redistribute
flux from real sources to noise peaks, producing spurious sources. 
This effect is generally only a problem for snapshot observations where {\em uv} 
coverage is poor. Although our {\em uv} coverage is very good, tests were performed to 
check the magnitude of the clean bias in our image.

We performed a clean bias check by injecting point sources into the {\em uv} data 
at random positions. The {\em uv} data were then cleaned to the same rms level 
as our original image, and source peak fluxes after cleaning were compared 
to input values. The flux density distribution of injected sources was chosen 
to be similar to that of the real sources in the image : 40 at $5 \sigma$, 15 at 
$6 \sigma$, 15 at $7\sigma$, 15 at $8\sigma$, 15 at $9\sigma$, 10 at $10\sigma$,
3 at $20 \sigma$, 2 at $30 \sigma$, 1 at $50 \sigma$ and 1 at $100 \sigma$. This 
simulation was repeated 50 times to get reliable number statistics. 

\begin{figure}
\centering
\includegraphics[width = 8cm, angle=-90]{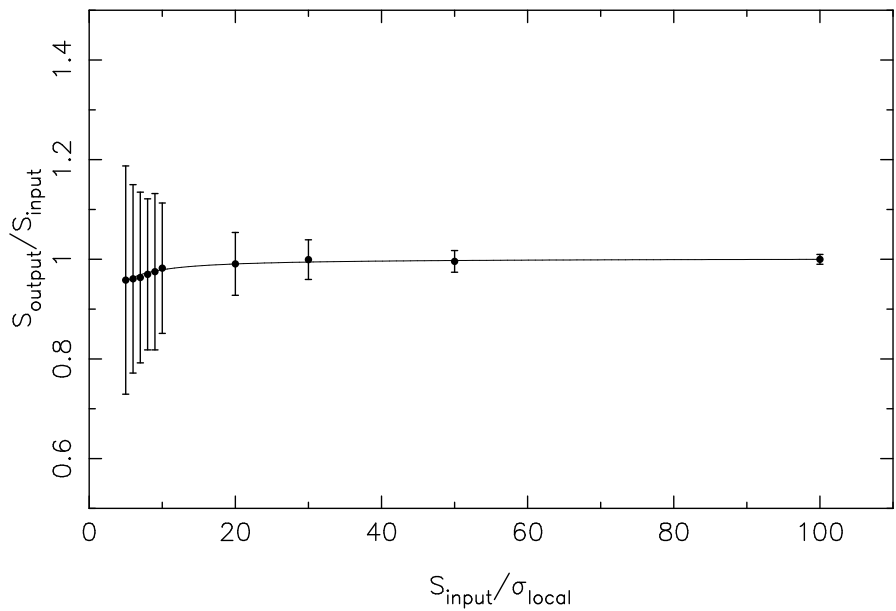}
\caption{The source flux measured after the cleaning process (S$_{\rm output}$) normalised to the 
true source flux (S$_{\rm input}$), as a function of the input source signal-to-noise 
(S$_{\rm input}/\sigma_{\rm local}$). Also shown is the best fit curve (see
Section \ref{cleanbias}), which is 
indistinguishable from the data.}
\label{fig:clnbias}
\end{figure}

The results from this process are shown in Figure \ref{fig:clnbias}. The average 
source flux measured after the cleaning process (S$_{\rm output}$) normalised to the 
true source flux (S$_{\rm input}$) is shown for the various values of input 
source signal-to-noise (S$_{\rm input}/\sigma_{\rm local}$). It is evident that 
clean bias only affects the faintest sources. There is a sudden increase in the clean 
bias at a flux density of about $10\sigma$, but even at the faintest bin ($5 \sigma$) the 
clean bias is only 4\%.

An analytical form of the clean bias effect was obtained by a least squares 
fit to the function
\[ S_{\rm output}/S_{\rm input} = a + b 
\frac{1}{S_{\rm input}/\sigma_{\rm local}} .\,\] 
The best fit  values for $a$ and $b$ are 1.00 and $-$0.23, respectively. The resulting best fit curve is 
plotted in Figure \ref{fig:clnbias}.

\subsection{Bandwidth Smearing}
\label{bandsmear}

Bandwidth smearing, the radio analogue of optical chromatic aberration, is a well-known effect 
caused by the finite width of the receiver channels. It reduces the 
peak flux density of a source, while correspondingly increasing the source 
size in the radial direction such that the total integrated flux density is 
conserved. The amount of smearing is proportional to the distance from the 
phase centre, and the bandwidth of the observations. 
In the simplest case of a Gaussian beam and passband, the bandwidth smearing 
can be described by \citep{condon98}:
\begin{equation}
\frac{S_{\rm peak}}{S_{\rm peak}^{O}} \approx \frac{1}{\sqrt{1 + \frac{2 \ln 2}{3} \left( \frac{\Delta \nu}{\nu} \frac {d}{\theta_b} \right)^2 }}
\label{eq:bandsmear}
\end{equation}
where the ratio $\frac{S_{\rm peak}}{S_{\rm peak}^{O}}$ represents the 
attenuation of the source peak flux density with respect to that of an 
unsmeared source, $\Delta \nu$ is the passband width, $\nu$ is the 
observing frequency, $d$ is the distance from the phase centre, and $\theta_b$ is the synthesized beam FWHM.

\begin{figure}
\centering
\includegraphics[width = 8cm]{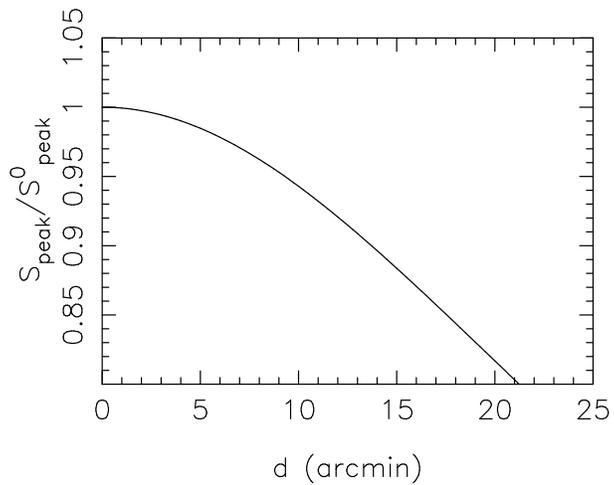}
\caption{Bandwidth smearing effect, as a function of distance from the phase centre.}
\label{fig:bandsmear}
\end{figure}

Although the observational bandwidth is 4 MHz, the {\em MIRIAD} task {\em
ATLOD} effectively doubles the channel bandwidth.  
So with the ATHDFS channel width ($\Delta \nu$) of 8 MHz and observing
frequency ($\nu$) of 1.4 GHz, Equation~\ref{eq:bandsmear} 
simplifies to :
\begin{equation}
\frac{S_{\rm peak}}{S_{\rm peak}^{O}} \approx \frac{1}{\sqrt{1 + \frac{2 \ln 2}{91875} \left( \frac {d}{\theta_b} \right)^2 }} \;.
\label{eq:bandsmear2}
\end{equation}
The bandwidth smearing as a function of distance from the centre is plotted 
in Figure \ref{fig:bandsmear}, calculated with the appropriate ATHDFS beamsize parameter.
The majority of our radio sources lie within the inner 10 $\times$ 10 
arcmin region, where the bandwidth smearing is less than 5\%. The 
attenuation increases rapidly however, and is $\sim 18$\% at a radial 
distance of 20 arcmin from the phase centre.

\section{Source Extraction}

\subsection{The Noise Map}
\label{sec:noiseim}
To investigate the noise characteristics of our 1.4 GHz image we constructed a noise map. 
The noise map which contains the pixel by pixel root mean square (rms) 
noise distribution was made using the SExtractor package \citep{bertin96}.
SExtractor initially estimates the local background in each mesh from the pixel data, and  
the local background histogram is clipped iteratively until convergence at 
$\pm3\sigma$ around its median. The choice of mesh size is very important. When it is too 
small the background will be overestimated due to the presence of real sources. 
When it is too large, any small scale variations of the background are 
washed out. 

\begin{figure}
\centering
\includegraphics[width = 8cm]{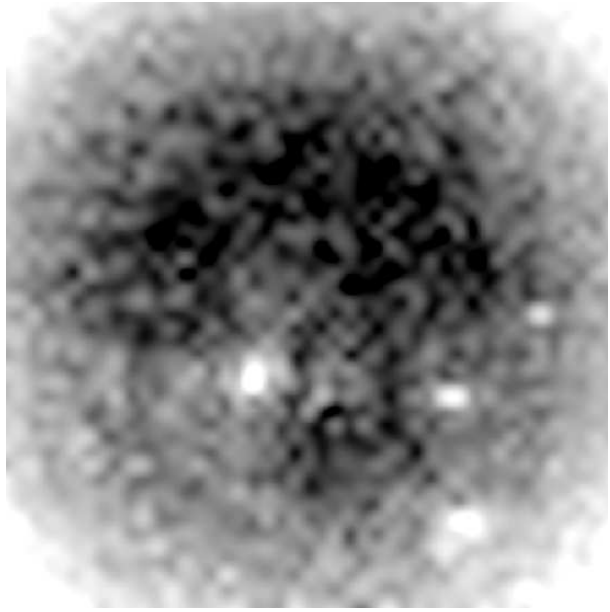}
\caption{A grey scale of the noise map obtained by SExtractor. The image is 40
$\times$ 40 arcmin, and darker regions indicate lower noise.}
\label{noisemap}
\end{figure}

\begin{figure}
\centering
\includegraphics[width = 8cm]{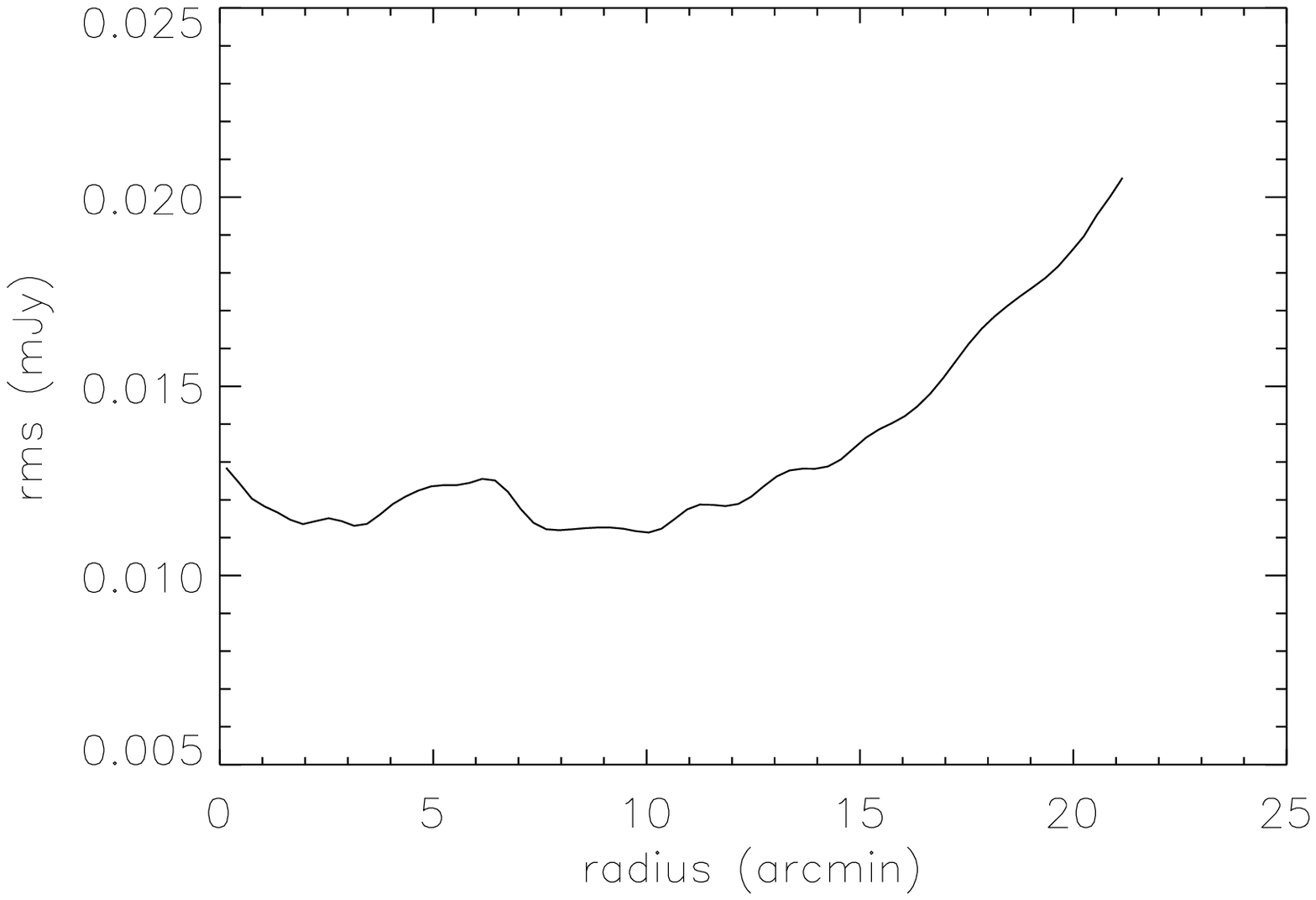}
\includegraphics[width = 8cm]{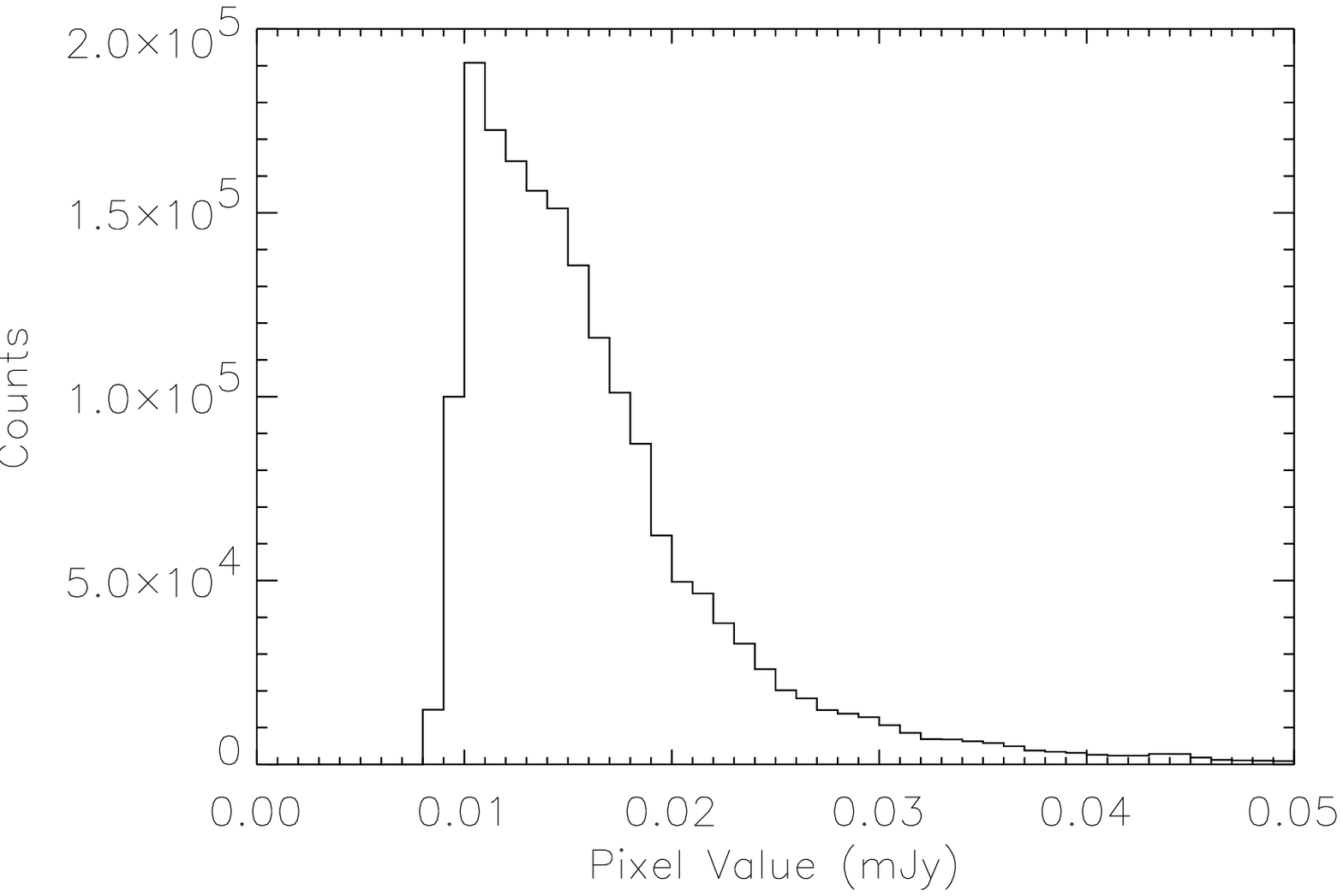}
\caption{Left panel: Noise (radially averaged) as a function of radial distance for the SExtractor 
noise map. The broad peak around 5-7 arcmin is due to the noise region around the bright 155 mJy source (ATHDFS\_J223355.6$-$604315). Right panel: Distribution of the pixel values of the SExtractor noise map.}
\label{noiseprops}
\end{figure}

We ran SExtractor with a mesh size set to 30 $\times$ 30 pixels, 
approximately 8 $\times$ 8 beams. A grey scale image of the SExtractor noise map is shown in 
Figure \ref{noisemap}. As expected, the noise is lowest in the centre and increases with 
radial distance. The left panel of Figure \ref{noiseprops} shows the average noise as a 
function of distance from the centre. This increase in noise is roughly a parabolic shape 
due to the primary beam attenuation. A histogram of the noise map pixel values is shown 
in the right panel of Figure \ref{noiseprops}. This distribution peaks at a value of 
10 $\mu$Jy, but has a large wing at high noise values ($\gtrsim15 \mu$Jy) due to the radially 
increasing noise.

There are also regions of increased noise (white pixels) around bright sources 
which are clearly evident in the grey scale image (Figure \ref{noisemap}). The increased noise 
around these sources is due to dynamic range effects, as discussed in Paper I. The most 
affected region is noticeable as a high noise bump at approximately 6~arcmin from the image centre 
in the left panel of Figure \ref{noiseprops}. This corresponds to the area surrounding the 
brightest source ($S_{\rm 1.4\,GHz}$ = 155 mJy). The noise in this region is up to 3 times 
that of the surrounding unaffected region.

\begin{figure}
\centering
\includegraphics[width = 7cm, angle=-90]{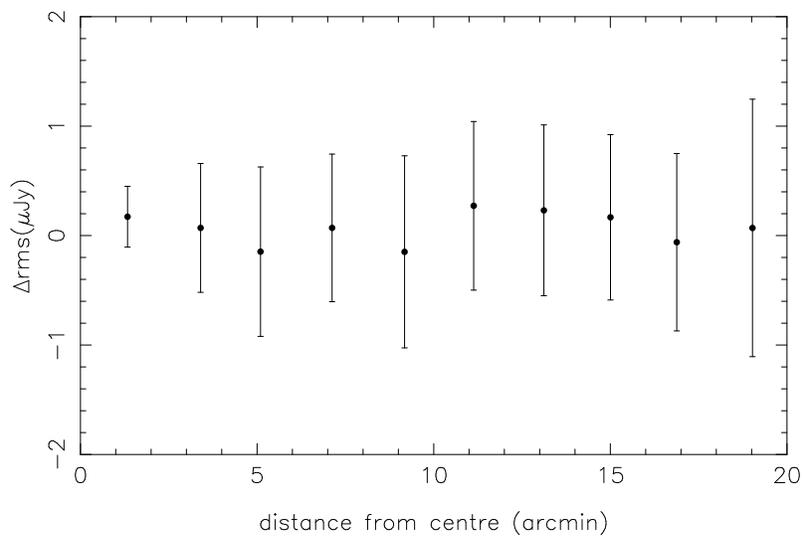}
\caption{Mean difference between directly measured rms value and the 
corresponding value in the SExtractor noise map as a function of radial
distance.}
\label{noisecompare}
\end{figure}

Although SExtractor was developed for the analysis of optical data, the noise image obtained 
with it has been found to be reliable for radio images \citep{bondi03}. We 
confirmed the accuracy of the SExtractor noise map by comparing pixel values against 
rms values directly measured from our radio image. The directly measured values were 
obtained by selecting a set of random positions in the 1.4 GHz image and calculating the rms 
in the surrounding $30 \times 30$ pixel box. Each box was examined for any sources above 
3$\sigma_{\rm local}$ before inclusion in the analysis to ensure that flux from real sources 
does not effect the statistics. Figure \ref{noisecompare} shows the mean and standard deviation 
of the difference between the SExtractor pixel values and the directly measured values as a 
function of distance from the centre. The difference is very close to zero, and there is no 
systematic effect as a function of radial distance. We therefore concluded that 
the SExtractor noise map is reliable and used it to perform the source extraction.

\subsection{Source Detection and Fitting}
\label{srcext}

We determined a sensible maximum radial distance for the source cataloging 
taking into account the following factors :\\
\indent i) the radial noise distribution, \\
\indent ii) the primary beam attenuation, and \\
\indent iii) the bandwidth smearing effect.\\ 
These three factors determine the efficacy of detecting a source with a particular 
flux density at different radial distances from the centre. The primary beam response 
at a radial distance of 20 arcmin is approximately 39\%, resulting in an increased noise of 
$\sim$20 $\mu$Jy, compared to $\sim$10 $\mu$Jy at the image centre. Furthermore, the bandwidth 
smearing effect is 18\% at this distance and rapidly degrading, as discussed in
Section~\ref{bandsmear}. On this basis we choose to catalogue to a maximum distance of 20 
arcmin from the image centre.

To begin the source extraction we divided the original 1.4 GHz map by the 
noise map generated by SExtractor, obtaining a `signal-to-noise' 
map.  SExtractor is useful for obtaining signal-to-noise maps, but it has not been reliably used for extracting sources from radio images. The commonly used routines are SFIND \citep{hopkins02}and {\it IMSAD}. Although SFIND may result in fewer false detections \citep{hopkins02}, there are no systematic differences in the source parameters found by either routine \citep{schinnerer04}. 
We used the {\it MIRIAD} task {\it IMSAD} to derive a preliminary 
list of source `islands'. This task searches for `islands' of pixels above a
specified cutoff, and attempts to fit gaussian components to the `islands'. 
We executed {\it IMSAD} on the `signal-to-noise' map to detect all islands 
with a peak flux density level of $4\sigma$.
This resulted in a list of 688 source `islands' for further investigation. 

Each source `island' found by {\it IMSAD} was examined and refitted with an elliptical 
Gaussian to derive source flux densities and sizes. This refitting procedure was performed on the `real' image (not the `signal-to-noise' map). All sources and fit parameters were 
visually inspected to check for obvious failures and poor fits that need further analysis. 
Following \cite{prandoni00b}, a reference peak value was derived from a second degree 
interpolation of the source ({\it MIRIAD} task {\it MAXFIT}). If the difference 
between the fitted peak and reference peak was less than 20\% of the reference value and 
the fitted position was inside the 0.9$S_{\rm peak}$ flux density contour,
then these sources were assigned a fit quality flag of 1. Gaussian 
integrated fluxes were also compared to the ones directly measured by summing the pixels 
greater than $3\sigma$ in the source area. In most cases the Gaussian fit provided 
good values for the position and peak flux densities but not integrated flux densities. 
These sources are given a quality flag of 2 in the catalogue.

There were some problematic cases which were classified as follows :\\
1) sources fit by {\em IMSAD} with a single Gaussian, but better described by 
two or more Gaussians ($\sim$3\%), and \\
2) non Gaussian sources not well fitted by a single Gaussian, i.e. fit exceeds 
tolerance criteria described above ($\sim$0.3\%), and \\
3) obviously spurious sources which correspond to artefacts or noise peaks 
($\sim$1\%). 

In the first case above, the {\it IMSAD} islands were split into two or three Gaussian 
components. The number of of successfully split islands are 19 in total 
(18 with two components, 1 with three components). The individual components
of these sources are listed in the catalogue. 

The reference positions and peak flux densities from {\it MAXFIT} were adopted for the two 
non Gaussian sources. The source position angle was determined by the direction along which 
it is most extended. The FWHM size of the axes was defined by the distance between 
two points on the $3\sigma$ contour parallel to (major) and perpendicular to (minor) 
this direction. These sources are flagged ``n'' in the catalogue.

\begin{figure}
\plottwo{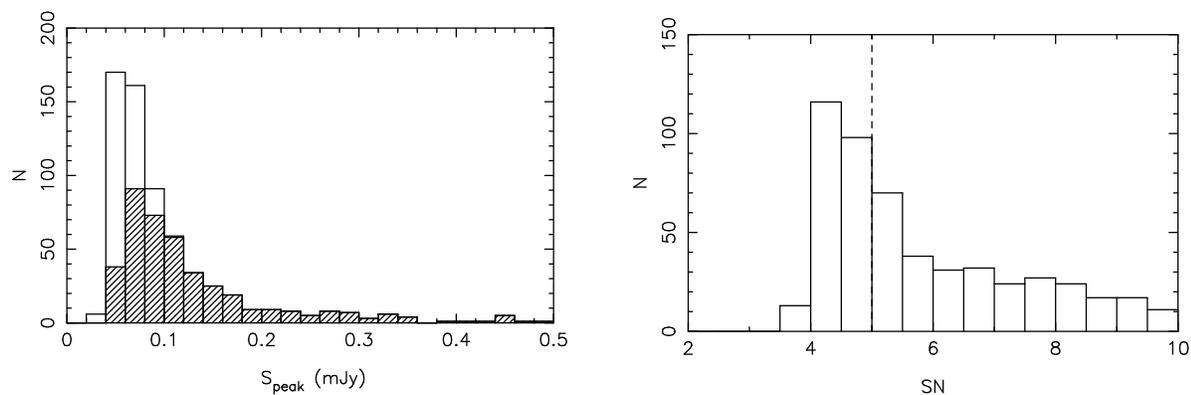}{f6b.eps}
\caption{Peak flux density (left panel) and signal-to-noise (right panel) distribution of the $4\sigma$ 
ATHDFS sample, extracted as described in Section \ref{srcext}. The peak flux
density distribution of the final $5\sigma$ catalogue is shown as the hatched
region in the left panel. The catalogue 5$\sigma$ cutoff is shown as a dashed
line in the right panel. }
\label{shist}
\end{figure}

\begin{figure}
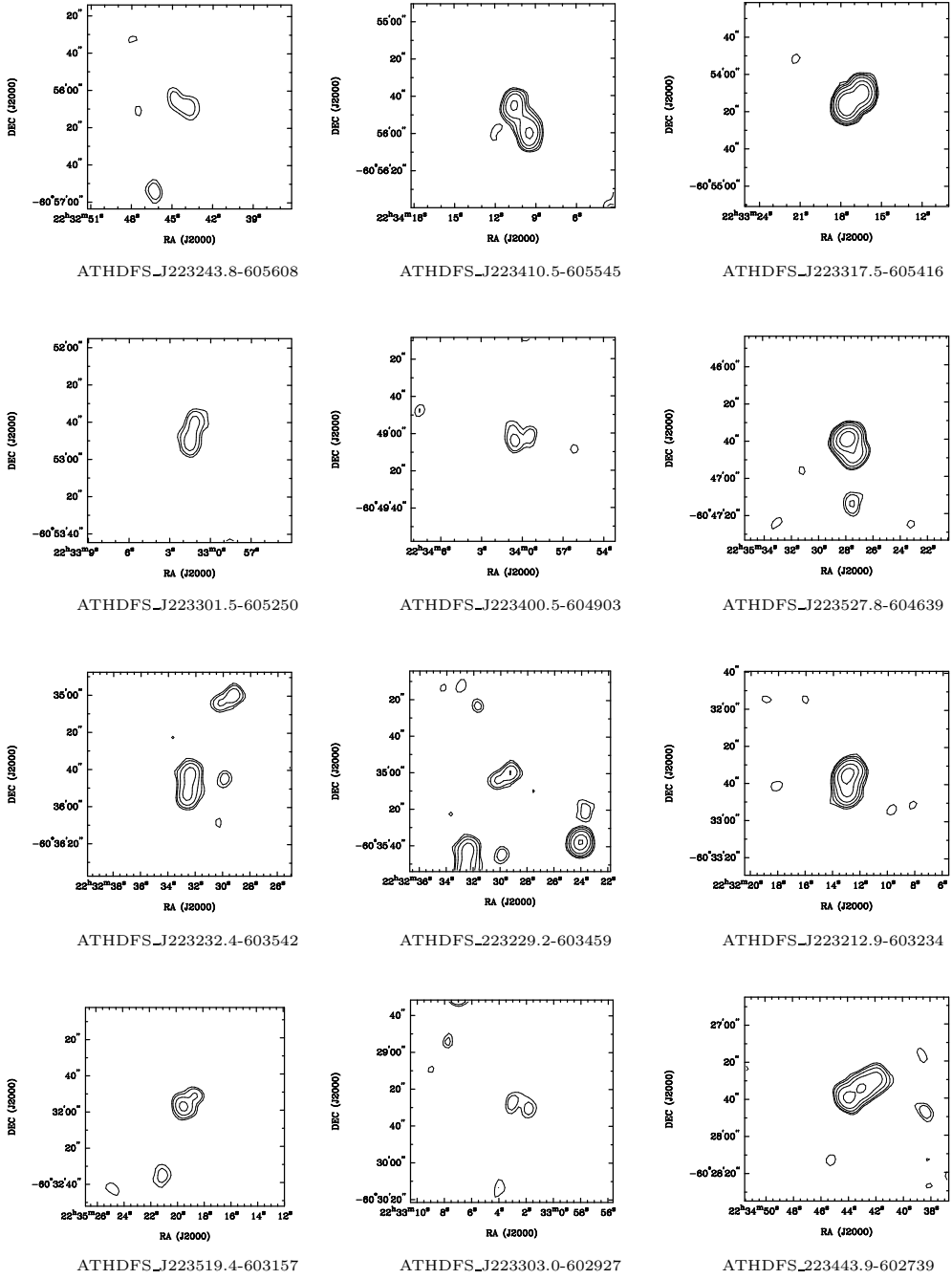

\begin{minipage}{19cm}

\includegraphics[width=4cm]{f7a.eps}

\vspace*{-3.44cm}{\hspace*{4.5cm}{\includegraphics[width=4cm]{f7b.eps}}}

\vspace*{-3.44cm}{\hspace*{9cm}{
\includegraphics[width=4cm]{f7c.eps}}}

\end{minipage}

\vspace*{-0.2cm}
{\tiny
\begin{tabbing}
\hspace*{0.9cm} \= \hspace*{4.4cm} \= \hspace*{4.4cm} \=
\hspace*{5.2cm}  \kill\> ATHDFS\_J223243.8-605608 \> ATHDFS\_J223410.5-605545 \>   ATHDFS\_J223317.5-605416 \\
\end{tabbing}
}

\begin{minipage}{19cm}

\includegraphics[width=4cm]{f7d.eps}

\vspace*{-3.44cm}{\hspace*{4.5cm}{\includegraphics[width=4cm]{f7e.eps}}}

\vspace*{-3.44cm}{\hspace*{9cm}{
\includegraphics[width=4cm]{f7f.eps}}}

\end{minipage}

\vspace*{-0.2cm}
{\tiny
\begin{tabbing}
\hspace*{0.9cm} \= \hspace*{4.4cm} \= \hspace*{4.4cm} \= \hspace*{4.4cm} \=
\hspace*{5.2cm}  \kill\>  ATHDFS\_J223301.5-605250 \>ATHDFS\_J223400.5-604903 \> ATHDFS\_J223527.8-604639  \\
\end{tabbing}
}

\begin{minipage}{19cm}

\includegraphics[width=4cm]{f7g.eps}

\vspace*{-3.44cm}{\hspace*{4.5cm}{\includegraphics[width=4cm]{f7h.eps}}}

\vspace*{-3.44cm}{\hspace*{9cm}{
\includegraphics[width=4cm]{f7i.eps}}}

\end{minipage}

\vspace*{-0.2cm}
{\tiny
\begin{tabbing}
\hspace*{0.9cm} \= \hspace*{4.4cm} \= \hspace*{4.4cm} \= \hspace*{4.4cm} \=
\hspace*{5.2cm}  \kill\> ATHDFS\_J223232.4-603542 \> ATHDFS\_223229.2-603459 \> ATHDFS\_J223212.9-603234 \\
\end{tabbing}
}

\begin{minipage}{19cm}

\includegraphics[width=4cm]{f7j.eps}

\vspace*{-3.44cm}{\hspace*{4.5cm}{\includegraphics[width=4cm]{f7k.eps}}}

\vspace*{-3.44cm}{\hspace*{9cm}{
\includegraphics[width=4cm]{f7l.eps}}}

\end{minipage}

\vspace*{-0.2cm}
{\tiny
\begin{tabbing}
\hspace*{0.9cm} \= \hspace*{4.4cm} \= \hspace*{4.4cm} \= \hspace*{4.4cm} \=
\hspace*{5.2cm}  \kill\>  ATHDFS\_J223519.4-603157 \> ATHDFS\_J223303.0-602927 \> ATHDFS\_223443.9-602739 \\
\end{tabbing}
}

\caption{Postage stamps, 1.8' $\times$ 1.8' in size, of the twelve 1.4 GHz 
sources fit by multiple Gaussians. The contour levels for each source are 
set at 3, 5, 10, 20, 50 and 100 $\sigma$.}
\label{multi_stamps}
\end{figure}

After accounting for the multiple components, non-Gaussian sources and noise artefacts, 
we have a list of 693 sources, or source components. Figure \ref{shist} shows the peak flux 
density and signal-to-noise distributions of the $4\sigma$ sample. The
spurious sources were low SN, and make up about 4\% of the 4 -- 5 $\sigma$ sources.
So to ensure we have a source catalogue of high reliability we apply a cut of
$5\sigma$, which includes 12 multiple component sources. 
There are clear signs of physical association between the individual components of these 
sources, as shown by the postage stamps in Figure~\ref{multi_stamps}.
The final 5$\sigma$ catalogue has 466 sources. 

\subsection{Deconvolution}
\label{sec:decon}

The ratio of integrated flux to peak flux is a direct measure of the
extension of a radio source:
\begin{equation}
S_{int} / S_{peak} = \frac{\theta_{maj} \theta_{min}}{b_{maj} b_{min}} \;,
\label{eq:fluxtheta}
\end{equation}
where $\theta_{maj}$ and $\theta_{min}$ are FWHM of the source axes, and $b_{maj}$ and $b_{min}$ the 
FWHM of the synthesised beam axes. We therefore use this relation to determine whether our sources are 
resolved or unresolved (see e.g. \citealp{prandoni00b, bondi03}).

\begin{figure}
\centering
\includegraphics[width = 8cm, angle=-90]{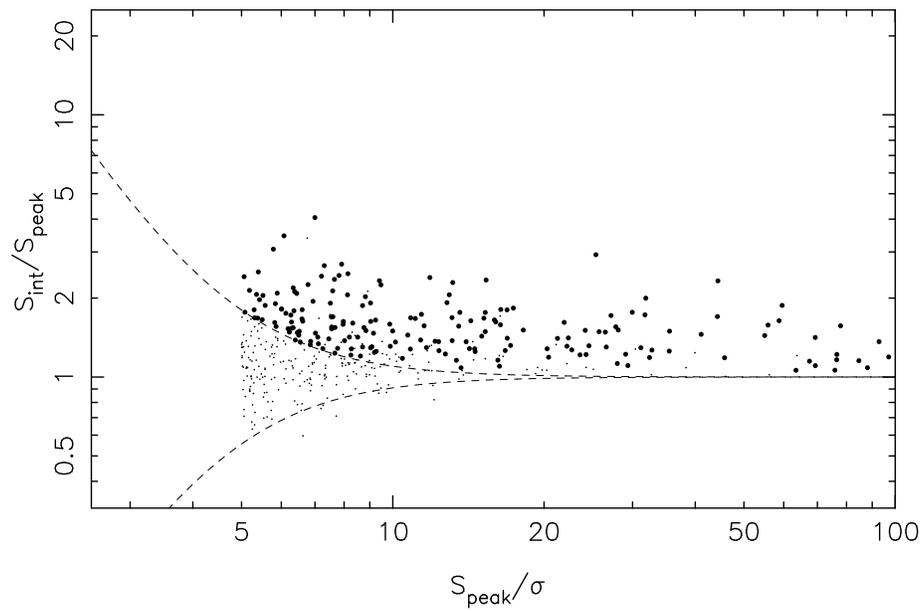}
\caption{Ratio of the integrated flux density (S$_{\rm int}$) to the peak flux density (S$_{\rm peak}$) as a 
function of source signal-to-noise ($S_{\rm peak}/\sigma$). The dashed lines show the upper and lower envelopes of the flux ratio 
distribution that contains 90\% of the unresolved sources ({\em small dots}). The
{\em large dots} indicate sources which are deconvolved successfully and considered resolved.}
\label{fig:deconvplot}
\end{figure}

In Figure \ref{fig:deconvplot} we plot the ratio of the total integrated flux density ($S_{\rm int}$) and peak 
flux density ($S_{\rm peak}$) as a function of signal-to-noise ($S_{\rm peak}/\sigma$) for all the sources 
in the catalogue. If the values of $S_{\rm int} < S_{\rm peak}$ are purely due to errors introduced 
by noise in our image, we can derive a criterion for extension by assuming that these statistical errors 
are also present for $S_{\rm int} > S_{\rm peak}$ sources. We determined the lower envelope in 
Figure \ref{fig:deconvplot} which contains 90\% of the sources $S_{\rm int} < S_{\rm peak}$ and mirrored 
it above the $S_{\rm int} = S_{\rm peak}$ line. The upper envelope 
is characterised by the~equation~:
\begin{equation}
S_{\rm int} / S_{\rm peak} = 1 + \left[ \frac{100}{(S_{\rm peak}/\sigma)^3} \right] \; .
\label{eq:deconv}
\end{equation}
Only sources lying above this
upper envelope and successfully deconvolved by {\it MIRIAD} are considered
resolved. From this analysis we conclude that 221/466 (47\%) of sources are
resolved. Deconvolved sizes are only given in the catalogue for resolved
sources, and other sources have deconvolved sizes set to zero.

\subsection{Multiple Component Sources}
\label{sec:multisources}
Radio sources can be made up of a nucleus with hot spots along, and at the end of, one or two jets. 
The individual components of a single source are often catalogued separately by Gaussian fitting 
routines. This will skew number statistics, so a 
method must be devised to identify multiple components as belonging to a single source. 

\begin{figure}
\centering
\includegraphics[width = 12cm]{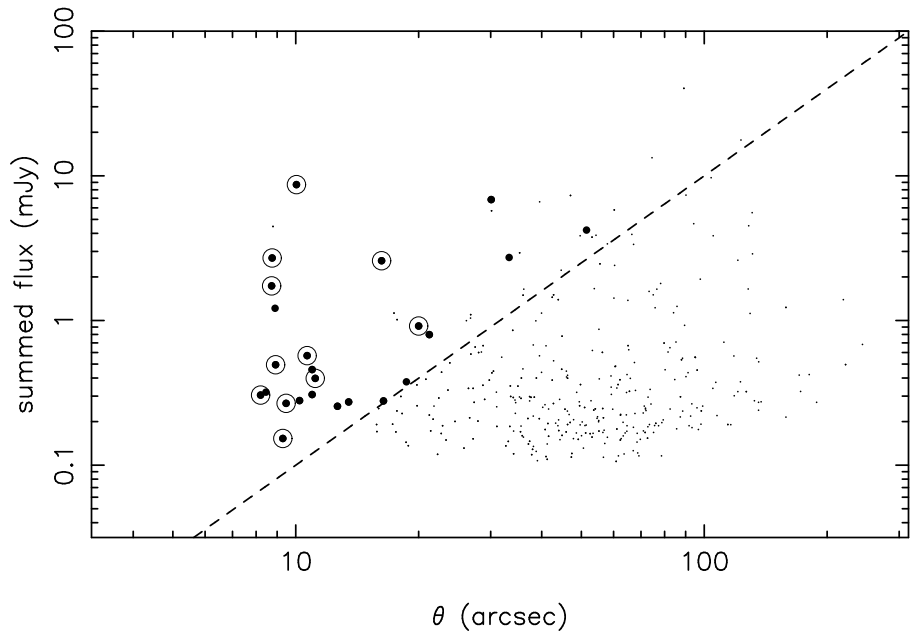}
\caption{Sum of the flux densities of nearest neighbour pairs plotted against
their separation. Source pairs which lie to the left of the line (see Section
\ref{sec:multisources} for details) are considered as possible double
sources. The source pairs with flux densities that differ by less than a
factor of 4 are shown as {\em large dots}. As a comparison, all sources already
well fit by 2 or more Gaussian components in Section \ref{srcext} 
have been outlined with a {\em large circle}.}
\label{nearest}
\end{figure}

Similar to the technique of \cite{maglio98}, we plot the sum of the fluxes of the components 
of each nearest neighbour pair versus their separation, as shown in Figure \ref{nearest}. The 
high density of points to the bottom right of the $\theta$-flux density plane is made up of the general
population of single component sources. The pairs with a separation of about 10 arcsec are most likely 
subcomponents of single sources. This is evident from Figure \ref{nearest} where the sources 
which were successfully split into multiple components in Section \ref{srcext}, shown as circled 
large dots, cluster around this separation. 

We apply a maximum separation that increases with the summed flux density, $S_{\rm total}$:
\begin{equation}
\theta = 100 \left( \frac{S_{\rm total} \rm{(mJy)}}{10} \right)^{0.5} {\rm arcsec} \;.
\label{summedfluxsep}
\end{equation}
This maximum separation is shown as a dashed line in Figure \ref{nearest}. By varying the 
allowed separation with summed flux, bright components may be considered a single source 
even at large separations, while faint sources are kept as single sources. 

As a further constraint we apply a flux ratio cut to the nearest neighbour pairs. Since  
the flux densities of real double sources are correlated this gives us another criterion 
to restrict the matched pairs to physically associated sources. We combine pairs only 
if their flux densities differ by a factor less than 4. The sources which meet this 
further requirement are marked in Figure \ref{nearest} by a large dot. For
sources which lie to the left of Equation \ref{summedfluxsep}, 20/23 pairs with
separations less than 20 arcsec pass the flux ratio cut, while only 3/21 pairs with
larger separations are passed. So this procedure seems 
to be successful in that almost all pairs with small separation are considered single 
sources, while very few distant pairs are passed.

From this analysis we identify an additional 11 possible multiple sources. All of these sources were 
visually inspected and only 5 pairs appear to be real double sources with lobe-lobe or 
core-lobe morphology. The postage stamps of these pairs are shown in Figure \ref{newmulti}. 
These sources are accepted as double sources and flagged in the catalogue. The other 6 pairs 
are shown in Figure \ref{notmulti}. They are rejected as double sources as they show no sign of 
physical association.

\subsection{The Catalogue}

The full $5\sigma$ catalogue is presented in Table \ref{catalogue}. 
A description of Table \ref{catalogue} is as follows. 

{\em Column (1)} --- Source ID.

{\em Column (2)} --- Source name. 

{\em Column (3)} --- Right Ascension in J2000.
 
{\em Column (4)} --- One sigma uncertainty of Right Ascension, in arcsec. This is calculated following \cite{condon97} and \cite{prandoni00b}.

{\em Column (5)} --- Declination in J2000.

{\em Column (6)} --- One sigma uncertainty of Declination, in arcsec. This is calculated following \cite{condon97} and \cite{prandoni00b}.

{\em Column (7)} --- Source 1.4 GHz peak flux density, in mJy. The values given 
here are not corrected for the systematic effects described in Section \ref{imageanalysis}
Appendix A describes how to obtain source peak flux densities corrected for the systematic effects.

{\em Column (8)} --- Source 1.4 GHz integrated flux density, in mJy.

{\em Column (9) and (10)} --- The {\em deconvolved} major and minor axes (FWHM)
                     of the source, $\Theta$, in arcsec. Zero values refer to unresolved sources 
(see Section \ref{sec:decon}).

{\em Column (11)} --- The {\em deconvolved} position angle (PA, measured from N through E)
                     of the source, in degrees. Zero values refer to unresolved sources 
(see Section \ref{sec:decon}).

{\em Column (12)} --- The signal-to-noise ratio of the detection,
                     calculated as {\em IMSAD} fitted peak/$\sigma_{\rm
                     local}$.

{\em Column (13)} --- Gaussian fit quality flag : ``1'' refers to very good
fits, ``2'' refers to poor integrated flux density, and ``n'' refers to non-Gaussian sources. See Section \ref{srcext} for more details.

{\em Column (14)} --- Notes on associated sources : components of a multiple source are indicated with the name of the other source(s) with which they are associated. See Section \ref{sec:multisources} for more details.

\section{Accuracy of Source Parameters}

\subsection{Internal Accuracy}
\label{sec:internalacc}

We performed Monte Carlo simulations to check the internal accuracy of our Gaussian fits 
to the ATHDFS radio sources. Firstly, a residual map was produced by removing all the 
sources detected above 5$\sigma$. Gaussian sources were injected into the residual image at random 
positions and the {\it MIRIAD} Gaussian fitting task {\it IMFIT} was used to extract fitted 
parameters. The injected point sources had peak flux densities between $5 \sigma$ and $100 
\sigma$. Fifty sources were injected at each peak flux density level per simulation (500 in total), and 
the simulation was repeated 20 times to get a sample size of 1000 sources per flux density level. 
By comparing the input to fitted output parameters we test the reliability of our source 
parameters, and obtain an estimate of the fit errors. 

In Figure \ref{fig:fitcheck1} the ratio of the fitted peak flux density 
(S$_{\rm output}$) to input peak flux density (S$_{\rm input}$) is shown as a function of the source 
signal-to-noise. There is an evident systematic effect (known as ``noise
bias'') in the fitted peak flux density for 
sources with signal-to-noise ratios less than about $10 \sigma$. Here the ratio of fitted to input peak 
flux density deviates from unity due to incompleteness at the lower signal-to-noise bins. If an injected 
source is coincident with a noise dip, its peak flux density is either under-estimated or it falls below the 
source detection threshold. This biases the fitted flux density to higher
values because only sources that fall on noise peaks, and therefore have extra
flux, are detected and measured.

Figure \ref{fig:fitcheck23} shows the ratio of output to input source sizes
for both major and minor axes. 
There maybe a marginal systematic over-estimation of the source major axis for
sources with signal-to-noise less than about $20 \sigma$. Although the output major axis 
sizes are increasingly over-estimated from $20 \sigma$ to $5 \sigma$, the
over-estimation is less than the standard deviation in the measurements. 
In contrast there is no systematic effect present on the measurement of 
the minor axis.

The accuracy of source positions was also checked in the simulations. The mean difference between 
output and input positions ($\Delta\alpha$ and $\Delta\delta$) are plotted in Figure \ref{fig:fitcheck45} 
as a function of source signal-to-noise. No systematics are present, and we notice that rms values for 
the two axes ($\alpha$ and $\delta$) are similar. The positional
accuracy of our catalogue is therefore derived as 1.1 arcsec at the limit of
our survey (5$\sigma$), improving to 0.6 arcsec at 10$\sigma$ and 0.1 arcsec at 100$\sigma$.

\subsection{Absolute Accuracy}

Ideally, our survey would be checked against an independent radio survey to
determine the absolute accuracy of the source parameters. Unfortunately there are no radio surveys
in the region which reach the required depth. The Parkes-MIT-NRAO (PMN) survey
\citep{griffith93} has a flux density limit of only 30 mJy at 4.85 GHz, while the Sydney
University Molonglo Sky Survey (SUMSS, \citealp{mauch03}) reaches a peak flux density 
limit of 6 mJy at 843 MHz. The SUMSS source positions are not useful
as a comparison because the SUMSS resolution is about 45 arcsec and source
positional uncertainties are up to 10 arcsec, which are much greater than the
ATHDFS internal positional uncertainties. Flux density comparisons are not
possible since these two surveys are at different frequencies and there is no
apriori knowledge of the source SEDs. 

The absolute positional accuracy of our survey will depend on how well the
positions of the phase calibrators are known. Calibrator PKS B2333-528 has
been observed by VLBI and its position in the FK5 reference system 
is known to within 0.005 arcsec in each coordinate \citep{ma98}. The position
of the second calibrator, PKS B2205-636, is known to within 0.1 -- 0.25
arcsec (ATCA calibrator manual, {\tt http://www.narrabri.atnf.csiro.au/calibrators/}). Thus we expect the absolute
accuracy of our source positions to be 0.25 arcsec. 
This absolute uncertainty in position is not included in the catalogue (Table
\ref{catalogue}), and needs to be added in quadrature to the source positional
uncertainty. This affects mainly the brighter sources, as the 
positional uncertainty of sources with signal-to-noise less 
than 10 are dominated by the internal accuracy of our survey. 

\section{Survey Completeness}

In the ATHDFS, sources were included in the catalogue if the measured peak flux density
exceeded five times the local noise. This can lead to incompleteness at faint
flux density levels, which has to be taken into account when deriving source
counts. In this Section we discuss the various possible sources of 
incompleteness in the source catalogue.

\subsection{Visibility Area}

Any source extraction method that relies on the ratio of peak flux density 
to the local noise will be affected by the variation of noise in the
image. The area over which a source of a given flux density can be
detected, also known as the visibility area, depends on the source peak flux
density and the homogeneity of the noise distribution. The noise in our image
is 11.0 $\mu$Jy at the centre and is a function of the primary beam response,
increasing towards the edges of the field, as quantified in Section \ref{sec:noiseim}. 
Local variation in image noise is also caused by bright 
sources, which can increase the noise in its vicinity by up to 3 times that of an 
unaffected region (see Section \ref{sec:noiseim} for details).

The visibility area of the ATHDFS 1.4 GHz image is shown in Figure
\ref{visfig}. As expected, the fraction of the total area over which a 
source of a given peak flux density can be detected increases 
between 0.05 and 0.1 mJy, and reaches 0.96 at 0.1 mJy. The rise in the
visibility function is not as rapid as in the ATESP \citep{prandoni01a} or
VIRMOS \citep{bondi03} surveys. This is because these surveys used a mosaicing
technique, which results in a more uniform noise distribution than
single-pointing surveys such as the ATHDFS.
The visibility area also shows that sources with a peak flux density greater 
than 0.1 mJy can be detected over more than 95\% of the whole image. The
visibility area reaches 100\% of the image at a peak flux density of 0.23 mJy. 

\subsection{Systematic Effects}

Two additional effects which could be responsible for partial incompleteness
in the source catalogue are bandwidth smearing and clean bias. These two effects have been extensively
discussed in Section \ref{imageanalysis}. Bandwidth smearing can lead to an underestimation of 
peak flux densities by up to $\sim$18\%, although the amount is 5\% or less for the
majority of our sources. The clean bias affects both peak and integrated flux
densities. It leads to an underestimation of flux densities by $\sim$5\% for the
faintest sources (5$\sigma_{\rm local}$), but has no effect on sources 
brighter than 10$\sigma_{\rm local}$. 

In Section \ref{imageanalysis} we proposed a formula (Equation A1) to correct the measured peak
and integrated flux densities for these two systematic effects. The correction
depends on the source signal-to-noise ratio as well as the source distance
from the image centre. However, peak flux densities are underestimated by 
$\sim$23\% {\em at most} due to the combined effects of both clean bias and
bandwidth smearing. Since the highest level of noise 
is $\sim$50 $\mu$Jy, it follows that clean bias and bandwidth smearing do
not affect sources brighter than $5 \times 50 \mu{\rm Jy} / 0.77 \sim 0.32$ mJy.

\subsection{Resolution Bias}

Resolution bias is an effect in which weak extended sources may have peak flux densities that fall below the 
catalogued 5$\sigma$ limit, yet still have total integrated flux densities 
above the survey limit. To derive a source count that is complete in terms of
total flux density, the number of ``missing'' sources has to be
estimated. Given a maximum detectable angular size and knowledge of the
intrinsic source size distribution as a function of flux density, 
this so called resolution bias can be corrected. Here we follow the procedures of
\cite{prandoni01a} to correct the resolution bias.

Assuming a Gaussian beam, the maximum size ($\theta_{\rm max}$) a source of total flux
density $S_{\rm total}$ can have before it drops below the $5\sigma_{\rm
local}$ detection limit can be calculated from the equation:
\begin{equation}
\frac{S_{\rm total}}{5\sigma_{\rm local}} = \frac{\theta_{\rm max}^2}{b_{\rm
maj}b_{\rm min}} \, .
\label{thetamaxfn}
\end{equation}
In Figure \ref{thetalims} we plot the angular size ($\theta$) of the ATHDFS
sources as a function of the measured total flux density. The angular sizes
are defined as the geometric mean of the major and minor deconvolved axes of
the sources. We assume $S_{\rm total} = S_{\rm peak}$ for point
sources. We calculate $\theta_{\rm max}$ from Equation \ref{thetamaxfn} with a
full range of noise values from 5 to 30 $\mu$Jy. The distribution of pixel 
values in the noise image of Section \ref{sec:noiseim} is used to determine the relative
weight of each noise value and a weighted average $\theta_{\rm max}$ is 
calculated. Figure \ref{thetalims} shows that the angular sizes of the largest 
ATHDFS sources are in good agreement with this weighted average $\theta_{\rm
max}$ function.

As discussed in \ref{sec:decon}, the deconvolution efficiency depends on the 
signal-to-noise of a source. An estimate of the minimum angular size ($\theta_{\rm
min}$) that a source can have is derived from Equations \ref{eq:fluxtheta} and \ref{eq:deconv}:
\begin{equation}
\frac{S_{\rm total}}{S_{\rm peak}} = 1 + 
\left[ \frac{100}{(S_{\rm total}/\sigma_{\rm local})^3} \right] = \frac{\theta_{\rm min}^2}{b_{\rm
maj}b_{\rm min}}  
\end{equation}
Similar to the derivation of $\theta_{\rm max}$, a weighted average
$\theta_{\rm min}$ is calculated using the noise distribution of our
image. The resulting average $\theta_{\rm min}$ is plotted in Figure
\ref{thetalims} as a dashed line. We note that the $\theta_{\rm min}$
constraint is important at low flux density levels where $\theta_{\rm max}$
becomes unphysical ($\rightarrow$0). Also, the minimum deconvolved 
size is about 10'' at the detection limit, whereas sources as small as 
$\sim$1'' are reliably deconvolved at higher signal-to-noise.

Using the two constraints described above, an overall angular size upper
limit, $\theta_{\rm lim}$, as a function of flux density $S_{\rm total}$ is defined as:
\begin{equation}
\theta_{\rm lim} = {\rm max}(\theta_{\rm max},  \theta_{\rm min})
\label{thetalimeq}
\end{equation}
 The incompleteness of the ATHDFS survey due to $\theta_{\rm lim}$ can be
estimated with knowledge of the true source angular size distribution as a function
of flux density. We assume an exponential form for the integral angular size
distribution, $h(\theta)$ \citep{windhorst90}:
\begin{equation}
h(\theta) = {\rm e}^{-{\rm ln}2(\theta/\theta_{\rm med})^{0.62}}
\label{windeq1}
\end{equation}
with
\begin{equation}
\theta_{\rm med} = 2'' S_{\rm 1.4GHz}^{0.30} \;,
\label{windeq2}
\end{equation}
where $ S_{\rm 1.4GHz}$ is in mJy, and  $\theta_{\rm med}$ is the estimated median 
source size at a given flux density. Equations \ref{windeq1} and \ref{windeq2} 
together with Equation \ref{thetalimeq} allow
us to estimate the fraction of sources larger than the maximum detectable
size and therefore missed by our survey. The correction factor, $c$, is then simply
calculated as:
\begin{equation}
c = \frac{1}{1 - h(\theta)} \;.
\label{c}
\end{equation}
In Figure \ref{rescorrfig} we show both $h(\theta)$ (left panel) and $c$
(right panel). The resolution bias peaks at about 0.1 mJy, but decreases at   
lower flux densities due to the $\theta_{\rm min}$ constraint.

\section{The ATHDFS 1.4 GHz Source Counts}

In this Section, we construct the differential radio source counts from the $5\sigma$ ATHDFS
catalogue. In computing the source counts we have used the
integrated flux density for resolved sources and peak flux densities for point
sources.  The counts were calculated by dividing 
the number of sources in each flux density bin by the total survey area and
the bin width, and then normalising by the Euclidean slope of $S^{-2.5}$. The
reason for the normalisation is historical, relating to the early use of radio
source counts to test various cosmological models against the number of
sources that would be expected in a steady state Euclidean universe \citep{longair66, ryle68}. 

Table \ref{countstable} lists the final ATHDFS source counts, corrected for
the visibility area, clean bias, bandwidth smearing and resolution bias as 
discussed above. For each flux density bin, the mean flux 
density ($<$$S$$>$), the number of detected sources ($N$), the number of effective sources after applying the corrections ($N_{\rm eff}$), and the corrected 
normalised differential radio source count ($dN/dS$), are given. We show the
counts before and after corrections in Figure \ref{athdfscounts1}. We notice
that the corrections are most important in the flux density range 0.09 -- 0.3
mJy, where they can increase the source count by up to approximately 30\%. 

Sources classed as "multiple-component" (see Section  4.4) only 
make one contribution to the source count. At these flux density levels, the difference between incorporating "multiple-component" sources as single objects and deriving counts from the "full component" catalogue is negligible. If all components are used, then the most significant change is in the $<$$S$$>$=0.119 mJy bin, where there are 4 additional sources. For this bin the difference is $\sim$ 7\%, which is less than the Poisson error.

\subsection{Comparison To Other Deep Radio Source Counts}

The ATHDFS source counts are compared to those derived from other 1.4 GHz 
surveys in Figure \ref{athdfscounts2}. The source counts from the Faint Images
of the Radio Sky at Twenty centimetres survey (FIRST, \citealp{white97}) 
are shown, as well as counts from the ATESP survey \citep{prandoni01a}, 
Phoenix Deep Field (PDF, \citealp{hopkins03}) and Hubble Deep Field North 
(HDFN, \citealp{richards00}). The solid line in Figure \ref{athdfscounts2} is the linear
least squares sixth order polynomial fit to this compilation of source
counts. FIRST counts below 2.5 mJy, were removed from the fit since
FIRST is incomplete at this limit. PDF counts above 2.5 mJy and ATESP counts
above 76 mJy were also excluded from the fit since the sampling is sparse in 
the higher flux density bins of these surveys. The resulting polynomial fit is
given by: 
\begin{equation}
\log[(dN/dS)/(S^{-2.5})] = \sum_{i=0}^{6} a_i(\log[S/{\rm mJy}])^i,
\end{equation}
with $a_0=0.841$, $a_1=0.540$, $a_2=0.364$, $a_3=-0.063$, $a_4=-0.107$,
$a_5=0.052$ and $a_6=-0.007$. For comparison, a third order fit to source 
counts from 0.1 mJy to 10000 mJy by \citep{katgert88} and a sixth order fit 
by \citep{hopkins03} to the PDF and FIRST source counts between 0.06 mJy and 
1 mJy are shown as dashed and dotted lines, respectively. A third order
polynomial is insufficient to account for the curvature below 0.1 mJy
suggested by both the HDFN and PDF surveys. A fourth order polynomial 
does not adequately model the point of inflection around 0.5 mJy \citep{hopkins03}, 
while fifth order polynomials do not reproduce the concave curvature seen 
at about 1 Jy (e.g. \citealp{windhorst90, katgert88}). Hence the sixth order 
polynomial is necessary. The fitted polynomial has no physical basis, but is
useful in quantifying the differential radio source count.

There is a high level of consistency between the source counts derived
from both small and large area surveys at flux densities $S_{\rm 1.4 GHz} >$  0.5
mJy (Figures 19 and 20). The differential source counts diverge at fainter flux density 
levels ($S_{\rm 1.4 GHz} <$ 0.5 mJy)  The source count from
the HDFN has been previously noted as being particularly low because of inadequate
completeness corrections, as well as cosmic variance \citep{richards00, hopkins03}. In contrast, the
count from the ATHDFS lies {\it higher} than all other counts at $S_{\rm 1.4 GHz} \sim$ 
0.4 mJy, although within statistical agreement for $S_{\rm 1.4 GHz} <$ 0.3
mJy.  The excess is significant for the ATHDFS at 0.37 mJy $< S_{\rm 1.4 GHz} <$ 
0.47 mJy compared to the latest (2003) PDF results and the counts from
the VIRMOS survey. Over the 0.35 deg$^2$ area catalogued by the ATHDFS survey, we
have 22 sources in this flux density range, compared to 44 in the VIRMOS field (1 deg$^2$) and 150 in the latest PDF catalogue (4.5 deg$^2$). We attribute the variation in the ATHDFS as real given
the small area of the ATHDFS, with a probable over density of $\sim$ 3 -- 8
sources compared to the 'average' field. 

At the very faintest limits ($S_{\rm 1.4 GHz} <$ 0.1 mJy) it has been suggested
that the differential source count steepens towards lower flux density
limits \citep{hopkins03, richards00}. Although there is little real confirmation, as in
every case it relies on the faintest source count bin, this steepening
is expected as the 'flat' region of the count between 0.08 and 0.3 mJy
cannot continue infinitely because the implied surface density of radio
sources would exceed the number of host galaxies \citep{windhorst93}. The
differential count from the ATHDFS is in agreement with the general
trend around $S$ = 0.08 mJy, although the last data point at
0.059 mJy lies well above those from the HDFN and PDF. 
Thus we cannot confirm the down-turn in the differential radio source count, and will
have to wait for a deeper source catalogue to supply this data.

\subsection{Source Counts and Galaxy Evolution}
\label{evolution}

Source counts are important because they can be used to constrain the
evolution of the radio population. The source count above 10 mJy is dominated 
by giant radio galaxies and QSOs, which we group together in this discussion
as AGN. At sub-mJy levels (0.1 -- 1 mJy) there is an increasing number of 
blue galaxies with starforming spectral signatures. \cite{condon89}, \cite{rowan93}, \cite{hopkins98} and others
have concluded that the source count at these faintest levels requires two
populations, AGN and starforming galaxies, both of which undergo strong evolution. 

To derive a local luminosity function of starforming galaxies,
\cite{hopkins98} start with the IRAS 60$\micron$ luminosity function
\citep{saunders90} and convert that to radio using the well-known 
FIR-radio correlation \citep{helou93, yun01}. Assuming that luminosity evolves
as $(1 + z)^q$, \cite{hopkins98} found $q = 3.3 \pm 0.8$ best matches the radio source counts.

We perform a similar test to determine the rate of evolution of the local
starforming galaxies required to match the radio source counts. We use the local
radio luminosity function of starforming galaxies determined by
\cite{condon02} for the 4583 Uppsala Galaxy Catalogue (UGC,
\citealp{nilson73}) galaxies crossmatched with the NRAO VLA Sky Survey (NVSS, \citealp{condon98}). 
The UGC-NVSS local RLF is the most comprehensive to date, very local ($z <
0.02$), and probes the low luminosity end of starforming galaxies 
to $\log (L_{\rm 1.4 GHz}) = 18.8$ W Hz$^{-1}$. The UGC-NVSS local RLF, $\rho_m(L)$, 
is described by the local visibility function, $\phi \equiv L^{5/2} \rho(L)$, where
\[ \log (\phi)  =  \log[\rho_m(L)] + \frac{3}{2} \log[L] + 28.43 \; . \]
Fitting the visibility function directly, instead of the RLF,
results in a better fit because the $L^{5/2}$ weighting prevents the visibility function from 
steepening as fast as the luminosity function at high 
luminosities. \cite{condon02} approximate the visibility function with a
hyperbolic function:
\[ \log(\phi) = {\rm Y} - \left\{ {\rm B}^2 + \left[ \frac{\log(L) - 
{\rm X}}{\rm W}
\right]^2 \right\}^{1/2}  \; , \]
with Y = 3.06, B = 1.9, X = 22.35, and W = 0.67, for starforming galaxies.
For the AGN component we use the AGN counts from \cite{jackson04}.

The observed source count is shown in Figure \ref{count_evolve} along with several
models for the counts of the local starforming population. We have explored a
variety of evolution scenarios: pure luminosity evolution going as $(1 +
z)^q$, pure density evolution going as $(1 + z)^p$, and a no evolution model.
We confirm that two distinct populations, the starforming galaxies
and AGN, are required to account for the faint source counts. The counts 
are best fit by starforming galaxies undergoing $(1 + z)^{2.7}$ luminosity
evolution. This amount of evolution means the number of
starforming galaxies exceeds the number of AGN at about 0.25 mJy. The counts
rule out stronger luminosity evolution of $q \gtrsim 4$. 
A modest amount of density evolution is also allowed, but if it is 
large (e.g. $p \gtrsim 6$) the source counts are too high at the
$\mu$Jy level. A downturn in radio source counts is needed at these levels so
that the number of radio sources don't exceed the number of host galaxies \citep{windhorst93}.
Thus, the radio source counts rule out large amounts of density evolution in
the starforming galaxy population. 

Similar values for the evolution of star forming galaxies have been found by \cite{hopkins04} and  \cite{seymour04}, who find $q = 2.7$ and $q = 2.5$, respectively. These two studies use the local radio luminosity function of starforming galaxies from \cite{sadler02}. Despite different assumptions for the luminosity functions of both starforming and AGN galaxies, there is good agreement between our work and these previous studies. 

Redshift information is required to break the degeneracy between luminosity and density
evolution. Little density evolution is expected however. For example, \cite{hopkins04} jointly constrain the evolution  parameters $q$ and $p$ to find $q = 2.7$ and $p = 0.15$. We have obtained spectra of our radio sources which will allow us to explore galaxy evolution in detail. This further work will be discussed in a future paper. 

\section{Summary}

We have presented the extended observations of the Hubble Deep Field South with the Australia Telescope 
Compact Array at 1.4 GHz. The new 1.4 GHz image reaches a maximum sensitivity of $\sim$11 $\mu$Jy rms at 
a resolution of $\sim$6.5 arcsec. The clean bias and bandwidth smearing effects have been investigated 
and quantified for our 1.4 GHz image. A catalogue of all sources brighter than $5\sigma_{\rm local}$ 
within 20 arcmin of the image centre has been compiled, comprising 466 sources. 

The 1.4 differential radio source counts from the ATHDFS survey have been
presented. We have corrected the counts for survey completeness and systematic
effects, including visibility area, clean bias, bandwidth smearing and
resolution bias. We find that the source counts in the ATHDFS are in general
consistent with that from previous surveys, although the ATHDFS counts are
greater than that found in the the HDFN by a factor of 1.2 -- 2. Other deep
surveys have also found the HDFN counts to be low (e.g. Phoenix,
\citealp{hopkins03}; VIRMOS, \citealp{bondi03}). Thus, the HDFN counts are
probably underestimated, or the HDFN samples a relatively under-dense region of radio
sources. We find the radio source counts are best fit by the local 
starforming population with luminosity evolution of the
form $(1 + z)^{2.7}$, although small amounts of density evolution are not ruled
out. 

The ATHDFS survey has obtained radio images at 2.5, 5.2 and 8.7 GHz in
addition to 1.4 GHz. The next paper in the series will provide radio spectral
indices and a detailed analysis of the ATHDFS sources across this radio spectrum. 
The Hubble Deep Field South has also been the target of deep multi-colour 
optical photometry and spectroscopy. Future papers in this series will present
the optical identifications and spectroscopy of the ATHDFS radio sources. 

\acknowledgements
{The authors thank the referee for helpful comments that improved this paper. 
MTH is grateful to have had support from an ANU PhD Stipend Scholarship and
CSIRO-CNR bilateral travel funds. The Australia Telescope Compact Array is part of the Australia Telescope
which is funded by the Commonwealth of Australia for operation as a National Facility managed by CSIRO.}

\appendix
\section{Flux density corrections for systematic effects}

As discussed in Section \ref{imageanalysis}, the two systematics which 
affect the peak flux densities of the ATHDFS sources have been analysed. 
These two effects are the clean bias and bandwidth smearing. The flux 
densities given in the ATHDFS catalogue (Table 2) are not corrected for these effects.
Corrected flux densities can be obtained using the following formula :
\begin{equation}
S_{\rm corr} = \frac{S_{\rm meas}{\sqrt{1 + \frac{2 \ln 2}{91875} ( \frac {d}{\theta_b} )^2 }}}{1.00 - 0.23 (S_{\rm meas}/\sigma)^{-1}},
\label{correction}
\end{equation}
where $S_{\rm meas}$ is the flux density actually measured and catalogued, 
$S_{\rm meas}/\sigma$ is the catalogued source signal-to-noise, $d$ is the 
source distance from the pointing centre (RA = 22h 33m 37.6s, Dec = 
$-$60$^\circ$ 33' 29''), and $\theta_b$ is the synthesized beam FWHM (6.6'').
The term in the numerator of Equation \ref{correction} corrects for 
the bandwidth smearing, while the term in the denominator represents the clean bias correction. 
For integrated flux densities, the term in the numerator of Equation \ref{correction} should be set to 1.


\clearpage

\clearpage

\begin{figure}
\begin{minipage}{19cm}

\includegraphics[width=4cm]{f10a.eps}

\vspace*{-3.44cm}{\hspace*{4.5cm}{\includegraphics[width=4cm]{f10b.eps}}}

\vspace*{-3.44cm}{\hspace*{9cm}{
\includegraphics[width=4cm]{f10c.eps}}}

\end{minipage}

\vspace*{-0.2cm}
{\tiny
\begin{tabbing}
\hspace*{0.9cm} \= \hspace*{4.4cm} \= \hspace*{4.4cm} \= \hspace*{4.4cm} \=
\hspace*{5.2cm}  \kill\> ATHDFS\_J223203.0-605242 \> ATHDFS\_J223524.8-604918 \>   ATHDFS\_J223327.9-603304 \> \\
\end{tabbing}
}

\begin{minipage}{19cm}

\includegraphics[width=4cm]{f10d.eps}

\vspace*{-3.44cm}{\hspace*{4.5cm}{\includegraphics[width=4cm]{f10e.eps}}}

\end{minipage}

\vspace*{-0.2cm}
{\tiny
\begin{tabbing}
\hspace*{0.9cm} \= \hspace*{4.4cm} \= \hspace*{4.4cm} \= \hspace*{4.4cm} \=
\hspace*{5.2cm}  \kill\> ATHDFS\_J223404.0-603037 \> ATHDFS\_J223227.6-602719 \> \> \\
\end{tabbing}
}

\caption{Postage stamps, 1.8' $\times$ 1.8' in size, of the 5 source pairs which pass the 
single source test described in Section \ref{sec:multisources} and appear to be physically associated. 
The contour levels for each source are set at 3, 5, 10, 20, 50 and 100 $\sigma$.}
\label{newmulti}
\end{figure}

\begin{figure}
\begin{minipage}{19cm}

\includegraphics[width=4cm]{f11a.eps}

\vspace*{-3.44cm}{\hspace*{4.5cm}{\includegraphics[width=4cm]{f11b.eps}}}

\vspace*{-3.44cm}{\hspace*{9cm}{
\includegraphics[width=4cm]{f11c.eps}}}

\end{minipage}

\vspace*{-0.2cm}
{\tiny
\begin{tabbing}
\hspace*{0.9cm} \= \hspace*{4.4cm} \= \hspace*{4.4cm} \= \hspace*{4.4cm} \=
\hspace*{5.2cm}  \kill\> ATHDFS\_J223458.6-605225 \> ATHDFS\_J223414.3-604923 \>   ATHDFS\_J223404.8-603732 \> \\
\end{tabbing}
}

\begin{minipage}{19cm}

\includegraphics[width=4cm]{f11d.eps}

\vspace*{-3.44cm}{\hspace*{4.5cm}{\includegraphics[width=4cm]{f11e.eps}}}

\vspace*{-3.44cm}{\hspace*{9cm}{
\includegraphics[width=4cm]{f11f.eps}}}

\end{minipage}

\vspace*{-0.2cm}
{\tiny
\begin{tabbing}
\hspace*{0.9cm} \= \hspace*{4.4cm} \= \hspace*{4.4cm} \= \hspace*{4.4cm} \=
\hspace*{5.2cm}  \kill\> ATHDFS\_J223400.2-603653 \> ATHDFS\_J223229.8-603544 \> ATHDFS\_J223509.5-603257 \> \\
\end{tabbing}
}

\caption{Postage stamps, 1.8' $\times$ 1.8' in size, of the 6 source pairs which pass the 
single source test described in Section \ref{sec:multisources} but are not taken to be physically associated. 
The contour levels for each source are set at 3, 5, 10, 20, 50 and 100 $\sigma$.}
\label{notmulti}
\end{figure}

\begin{figure}
\centering
\includegraphics[width = 7cm]{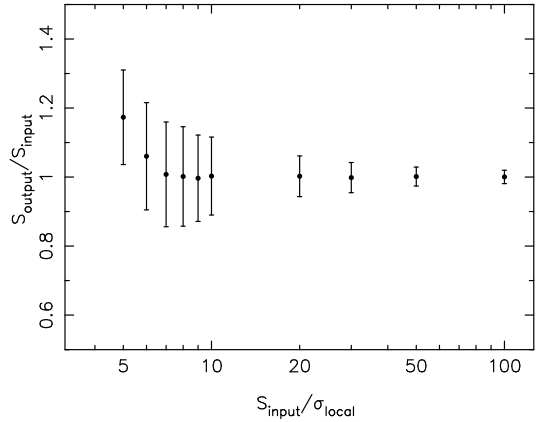}
\caption{The mean and standard deviation of output/input peak flux
densities for the injected sources, as a function of signal-to-noise.}
\label{fig:fitcheck1}
\end{figure}

\begin{figure}
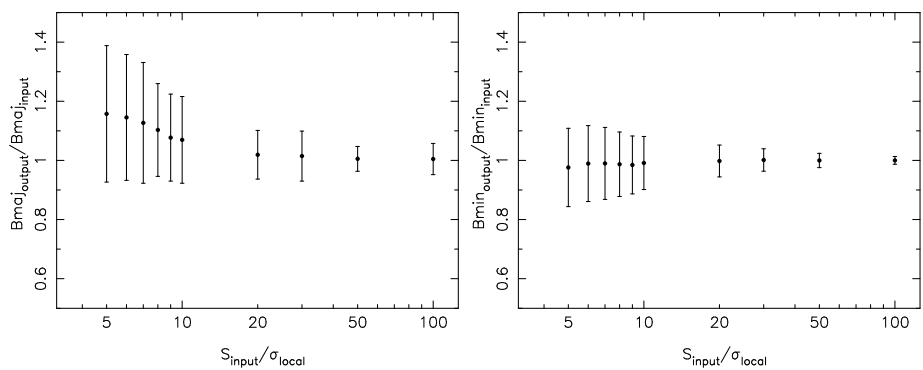

\centering
\includegraphics[width = 6cm]{f13a.eps}
\includegraphics[width = 6cm]{f13b.eps}
\caption{The mean and standard deviation of output/input axis FWHMs for the injected sources, 
as a function of signal-to-noise. Left panel : major axis. Right panel : minor axis. }
\label{fig:fitcheck23}
\end{figure}

\begin{figure}
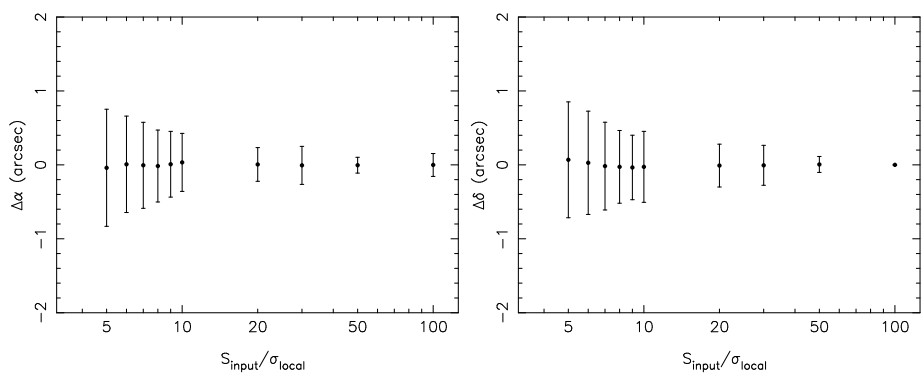

\centering
\includegraphics[width = 6cm]{f14a.eps}
\includegraphics[width = 6cm]{f14b.eps}
\caption{The mean and standard deviation for the difference in output position
to input positions of injected sources, as a function of signal-to-noise. Left panel : right ascension. Right panel : declination. }
\label{fig:fitcheck45}
\end{figure}

\clearpage

\begin{figure}
\centering
\includegraphics[width = 12cm]{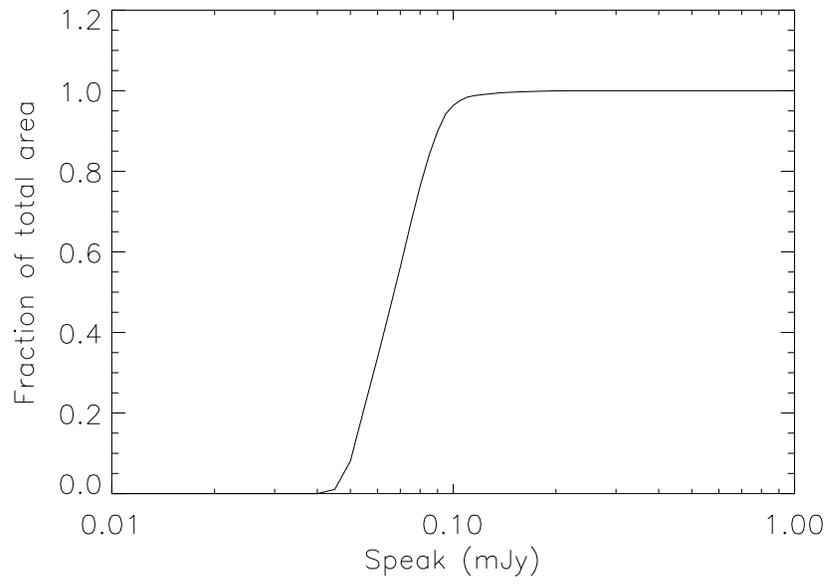}
\caption{Visibility area of the single-pointing ATHDFS 1.4 GHz image. Fraction of the total area over which a 
source of measured peak flux density $S_{\rm peak}$ could be detected. The
rise in visibility area is less steep than in \cite{prandoni01a} (c.f their
Figure 1.)}
\label{visfig}
\end{figure}

\begin{figure}
\centering
\includegraphics[width = 12cm]{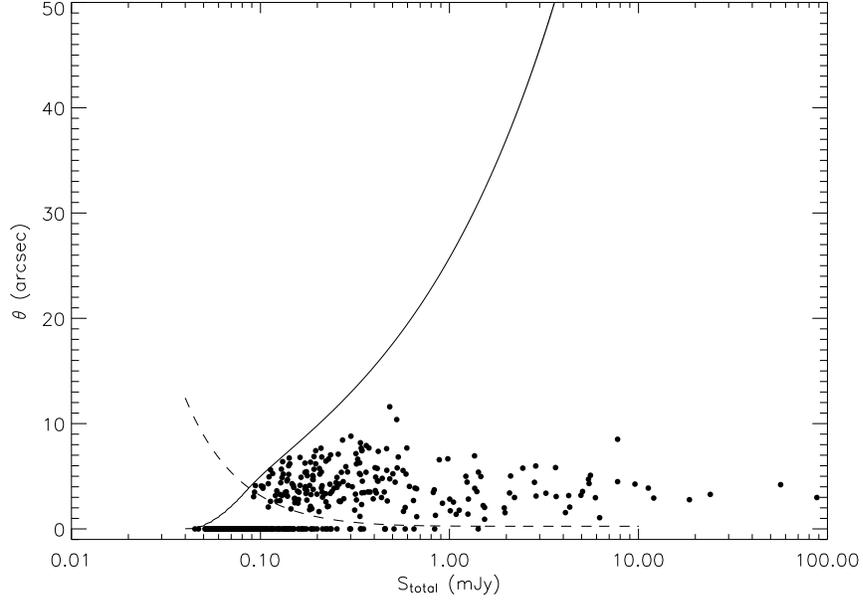}
\caption{Angular size (geometric mean of deconvolved axes) for the ATHDFS sources 
as a function of the measured to flux density. Unresolved sources have
deconvolved sizes, and therefore angular sizes, set to zero. 
The maximum size ($\theta_{\rm max}$) a source can have before dropping below the detection limit is shown as a solid line. 
Also drawn is the minimum angular size ($\theta_{\rm min}$) reliably deconvolved 
as a function of flux density (dashed line). }
\label{thetalims}
\end{figure}

\begin{figure}
\centering
\includegraphics[width = 7cm]{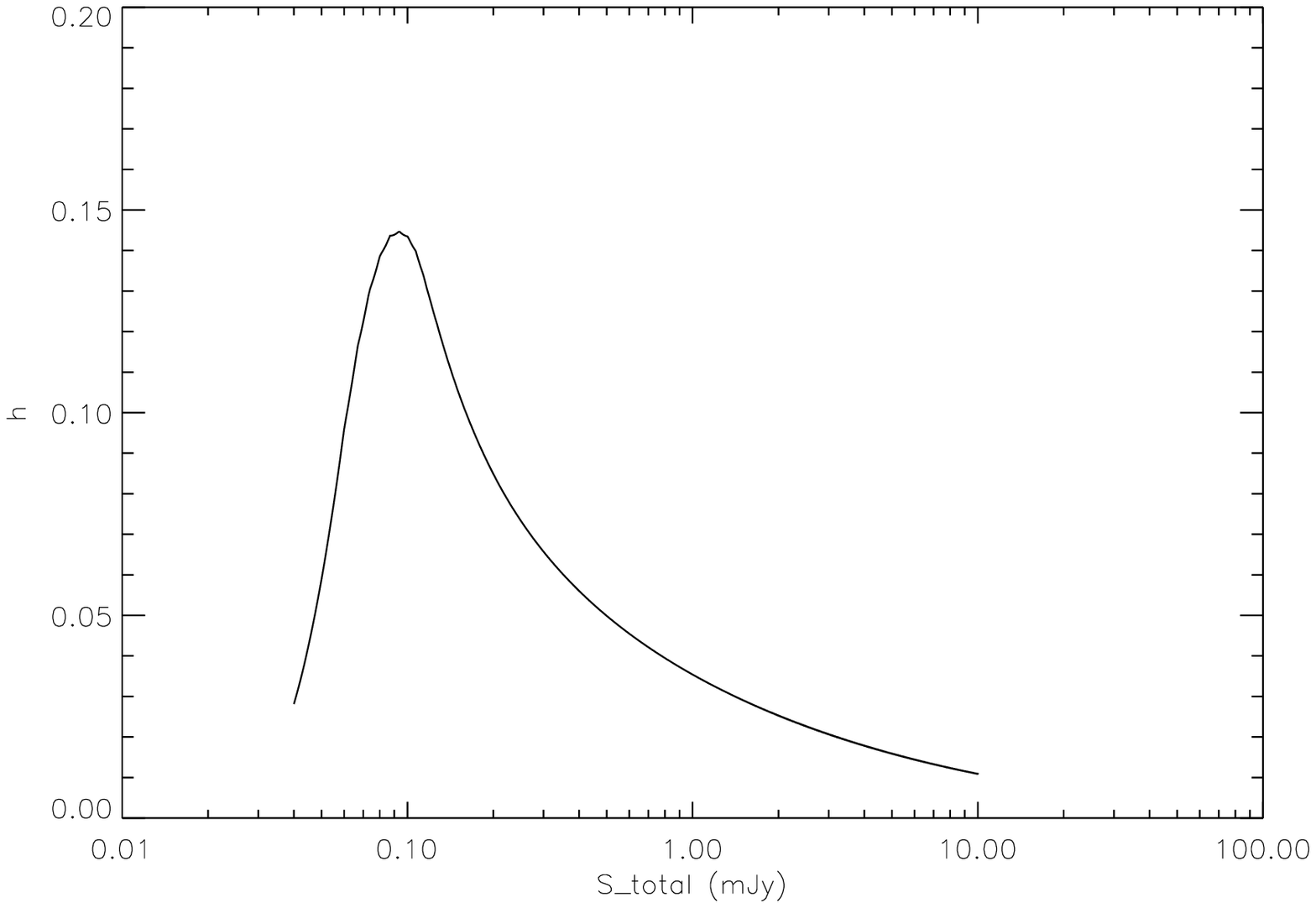}
\includegraphics[width = 7cm]{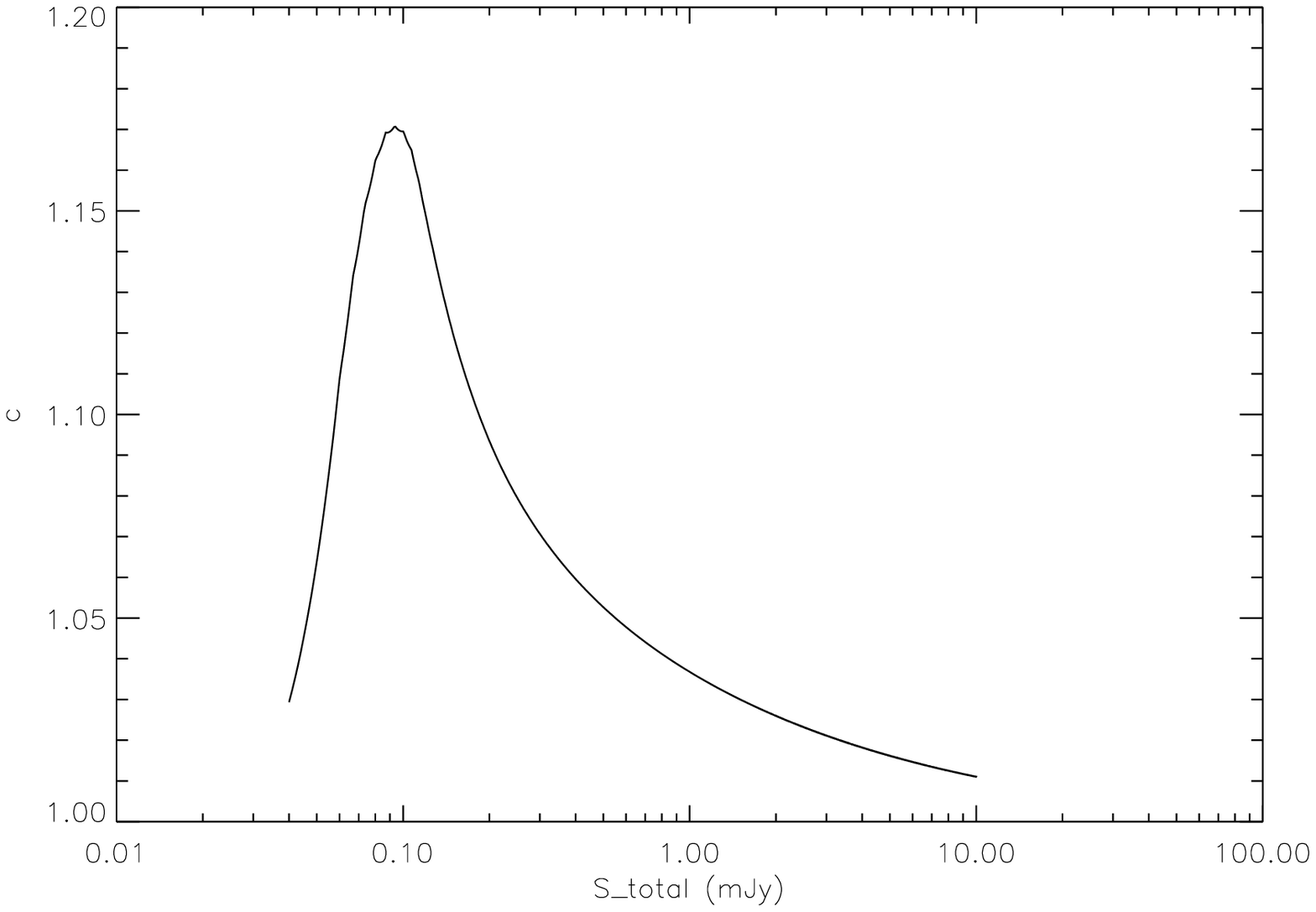}
\caption{Left panel: Fraction of sources with angular sizes greater than $\theta_{\rm lim}$
, $h(\theta_{\rm lim})$, as a function of flux density. Right panel: The resolution bias correction, 
$c = \frac{1}{1 - h(\theta_{\rm lim})}$.}
\label{rescorrfig}
\end{figure}

\begin{figure}
\centering
\includegraphics[width = 11cm]{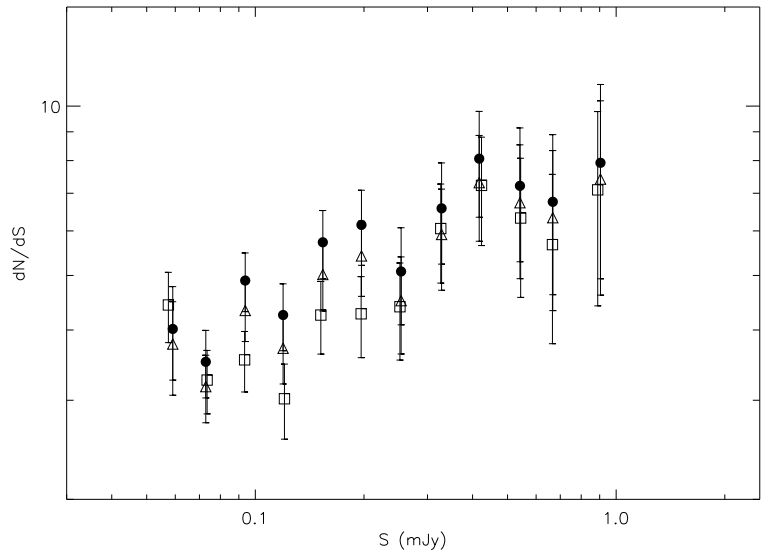}
\caption{The 1.4 GHz normalised differential radio source counts derived from the
ATHDFS catalogue. The source counts before correcting for 
clean bias, bandwidth smearing and resolution bias are shown as {\em
squares}. {\em Triangles} indicate the counts after correcting for clean bias
and bandwidth smearing, but not resolution bias. Source counts with full
corrections are shown as {\em circles}.}
\label{athdfscounts1}
\end{figure}

\begin{figure}
\centering
\includegraphics[width = 14cm]{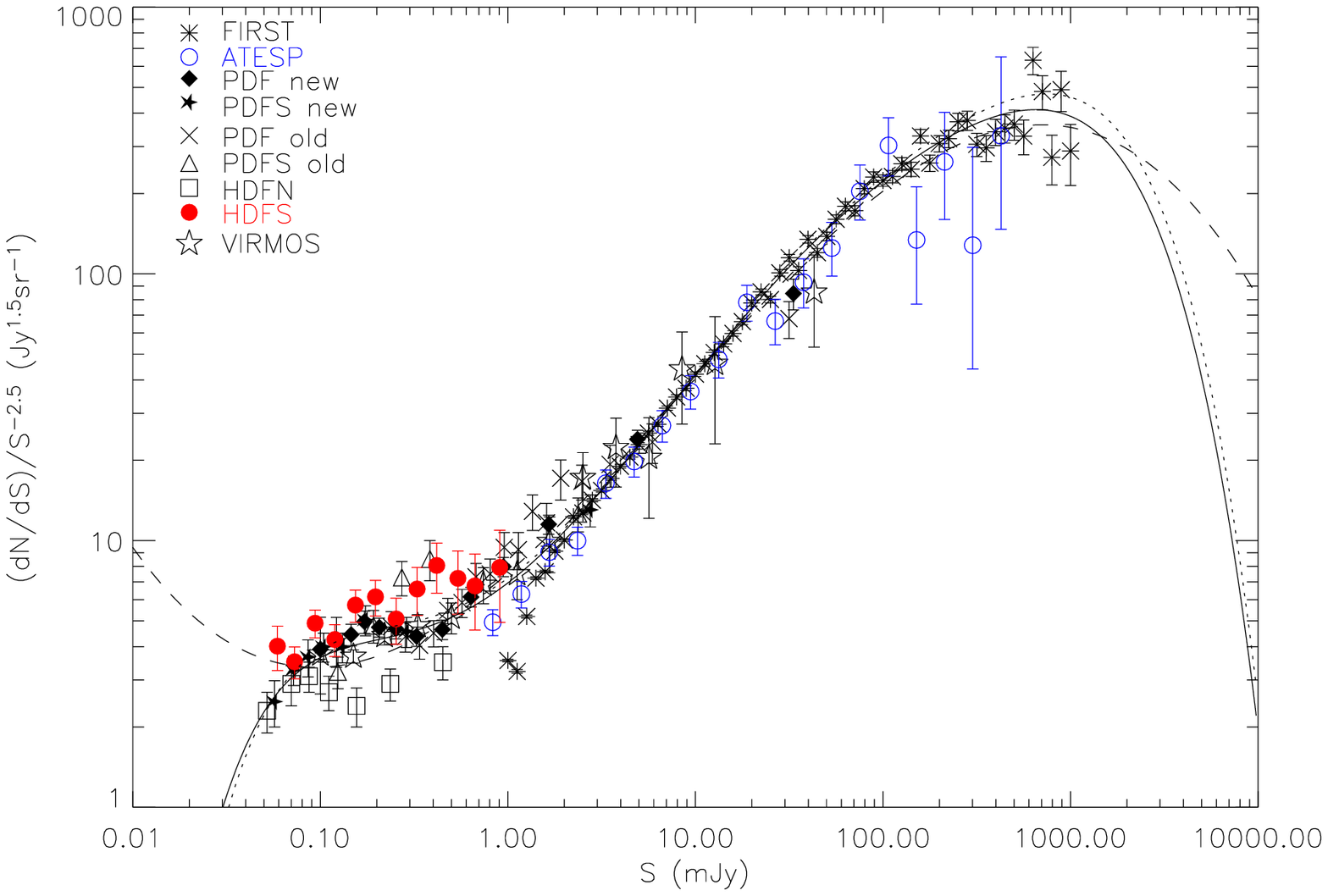}
\caption{Normalised 1.4 GHz differential radio source counts: FIRST
({\em asterisks}; \citealp{white97}), the original Phoenix Deep Field survey
({\em crosses} and {\em triangles}; \citealp{hopkins98}), the latest Phoenix Deep Field survey
({\em diamonds} and {\em filled stars}; \citealp{hopkins03}), the ATESP survey
({\em empty circles}; \citealp{prandoni01a}), the Hubble Deep Field North
({\em empty squares}; \citealp{richards00}), the ATHDFS ({\em filled circles}), and the VIRMOS survey
({\em empty stars}; \citealp{bondi03}). The solid line is a sixth order polynomial fit 
described in the text. The sixth order fit from \cite{hopkins03} 
(dotted line) and third order fit from \cite{katgert88} (dashed line) are
shown for comparison.}
\label{athdfscounts2}
\end{figure}

\begin{figure}
\centering
\includegraphics[width = 14cm]{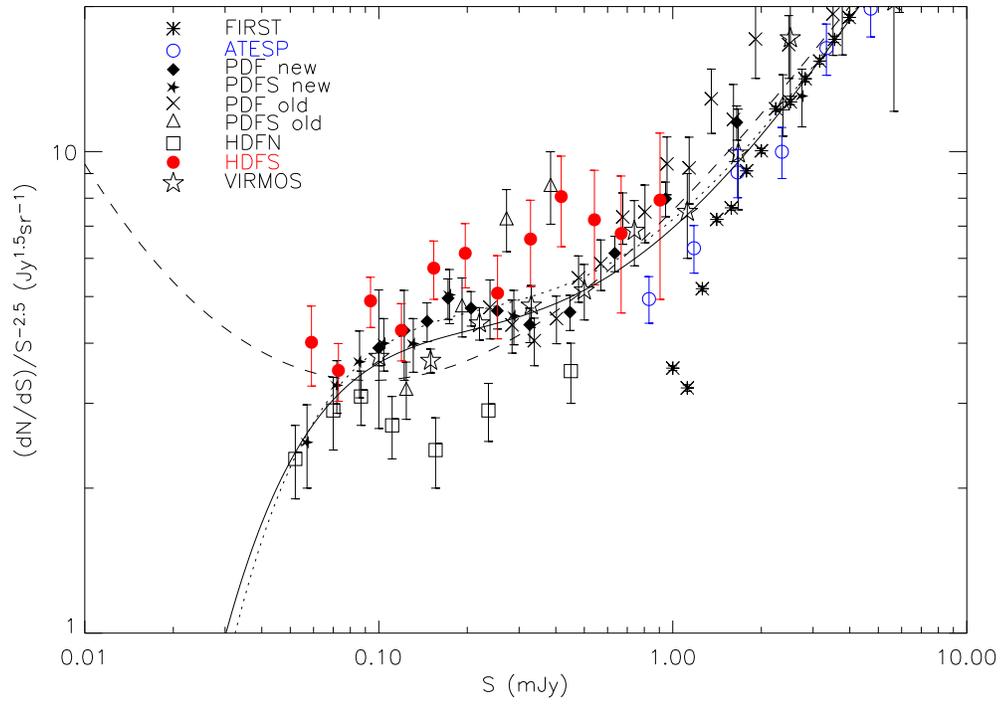}
\caption{Normalised 1.4 GHz differential radio source counts - zoomed
on the faint flux density end. Symbols and lines are the same as Figure \ref{athdfscounts2}.}
\label{athdfscounts3}
\end{figure}

\begin{figure}
\centering
\includegraphics[width = 14cm]{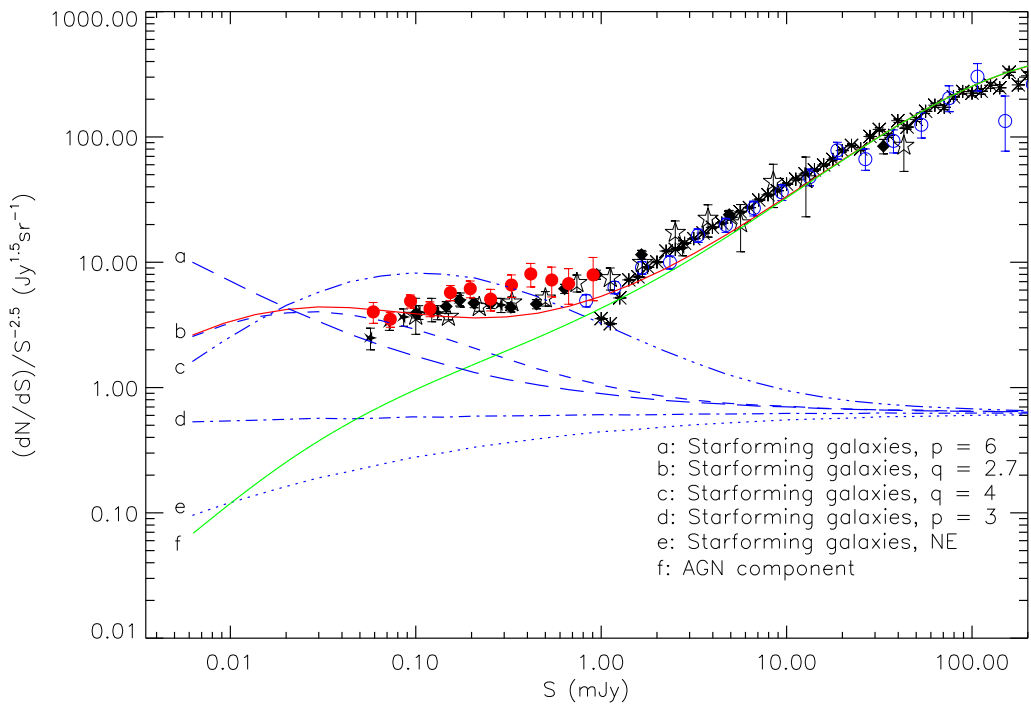}
\caption{Radio source count models, plotted with observed data. The
observed counts are as for Figure \ref{athdfscounts2}. See Section
\ref{evolution} for a description of the models. The solid line near 
``b'' is just model ``b'' added to the AGN component (line ``f'').}
\label{count_evolve}
\end{figure}


\begin{deluxetable}{llccccrrrrrrcc}
\rotate
\tablewidth{0pt}
\tabletypesize{\tiny}
\tablecaption{ATHDFS 1.4 GHz catalogue\label{catalog}}
\tablehead{\colhead{ID} & \colhead{source} & \colhead{RA} & \colhead{$\sigma\alpha$} & \colhead{Dec} & 
\colhead{$\sigma\delta$} & \colhead{$S_{\rm peak}$} &
\colhead{$S_{\rm int}$} & \colhead{$\theta_{maj}$} & \colhead{$\theta_{min}$} & \colhead{PA} & 
\colhead{SN$_{\rm local}$} & \colhead{Flag} & \colhead{Notes}}
\startdata
1  &  ATHDFS\_J223303.6$-$605751  &  22:33:03.62  &  0.12  &  $-$60:57:51.6  &  0.41  &  0.297  &  0.508  &  0.00  &  0.00  &  0.0  &  16.4  &  2 & \\
2  &  ATHDFS\_J223325.8$-$605729  &  22:33:25.88  &  0.55  &  $-$60:57:29.3  &  1.28  &  0.092  &  0.162  &  7.81  &  5.90  &  $-$5.3  &  5.1  &  2 & \\
3  &  ATHDFS\_J223255.6$-$605656  &  22:32:55.62  &  0.48  &  $-$60:56:56.5  &  0.76  &  0.159  &  0.375  &  9.35  &  6.32  &  $-$0.5  &  7.7  &  2 & \\
4  &  ATHDFS\_J223246.3$-$605654  &  22:32:46.30  &  0.28  &  $-$60:56:54.0  &  0.50  &  0.170  &  0.254  &  6.92  &  2.24  &  18.6  &  10.0  &  2 & \\
5  &  ATHDFS\_J223403.4$-$605640  &  22:34:03.43  &  0.54  &  $-$60:56:40.3  &  0.66  &  0.126  &  0.210  &  5.63  &  5.20  &  $-$7.8  &  6.6  &  2 & \\
6  &  ATHDFS\_J223243.8$-$605608  &  22:32:43.80  &  0.60  &  $-$60:56:08.8  &  0.62  &  0.149  &  0.363  &  8.55  &  7.33  &  57.1  &  7.8  & 1  & ATHDFS\_J223244.8$-$605603 \\
7  &  ATHDFS\_J223244.8$-$605603  &  22:32:44.89  &  0.43  &  $-$60:56:03.7  &  0.84  &  0.119  &  0.176  &  7.41  &  1.01  &  19.2  &  6.2  &  1 & ATHDFS\_J223243.8$-$605608 \\
8  &  ATHDFS\_J223410.5$-$605545  &  22:34:10.58  &  0.05  &  $-$60:55:45.4  &  0.09  &  1.284  &  2.106  &  7.28  &  3.47  &  $-$3.9  &  58.7  &  1 & ATHDFS\_J223409.5$-$605559 \\
\enddata
\label{catalogue}
\end{deluxetable}

\begin{deluxetable}{ccrrc}
\tablewidth{0pt}
\tablecaption{The ATHDFS 1.4 GHz source counts.\label{countstable}}
\tablehead{ \colhead{Range in $S$ (mJy)} & \colhead{$<$$S$$>$ (mJy)} & \colhead{$N$} & \colhead{$N_{\rm eff}$} 
& \colhead{$(dN/dS)/S^{-2.5}$ (${\rm Jy}^{1.5}{\rm sr}^{-1}$)} }
\startdata
0.050 -- 0.064 & 0.059 & 28 & 226.1 & 4.02 $\pm$ 0.76 \\
0.064 -- 0.082 & 0.073 & 53 & 149.8 & 3.51 $\pm$ 0.48 \\
0.082 -- 0.106 & 0.094 & 70 & 143.0 & 4.90 $\pm$ 0.59 \\
0.106 -- 0.136 & 0.119 & 54 & 87.06& 4.25 $\pm$ 0.58 \\
0.136 -- 0.174 & 0.154 & 52 & 79.85 & 5.73 $\pm$ 0.79 \\
0.174 -- 0.224 & 0.197 & 43 & 59.43 & 6.15 $\pm$ 0.94 \\
0.224 -- 0.287 & 0.253 & 26 & 33.49 & 5.08 $\pm$ 1.00 \\
0.287 -- 0.368 & 0.328 & 24 & 29.18 & 6.58 $\pm$ 1.34 \\
0.368 -- 0.473 & 0.417 & 22 & 25.18 & 8.06 $\pm$ 1.72 \\
0.473 -- 0.607 & 0.541 & 14 & 15.10 & 7.22 $\pm$ 1.93 \\
0.607 -- 0.779 & 0.667 & 10 & 10.75 & 6.76 $\pm$ 2.14 \\
0.779 -- 1.000 & 0.905 & 7  & 7.56 & 7.93 $\pm$ 3.00 \\
\enddata
\end{deluxetable}

\end{document}